\documentclass{aa}  

\usepackage{graphicx}
\usepackage{txfonts}
\usepackage{xcolor}
\usepackage{ragged2e}   % added MR
\usepackage{placeins}   % added MR

\usepackage[colorlinks, citecolor = blue]{hyperref}
\begin{document} 

\title{PRUSSIC~I - a JVLA survey of HCN, HCO$^+$, and HNC (1--0) emission in $z\sim3$ dusty galaxies\thanks{This project has been inspired by Yu Gao's seminal works on dense-gas tracers. Sadly, Yu Gao passed away during the review of this paper.}}
\subtitle{Low dense-gas fractions in high-redshift star-forming galaxies}

   \author{
        M. Rybak\inst{1,2}
        \and J. A. Hodge\inst{2}
        \and T. R. Greve\inst{3,4,5}
        \and D. Riechers\inst{6}
        \and I. Lamperti\inst{7}
        \and J. van Marrewijk\inst{2,8}
        \and F. Walter\inst{9}
        \and J. Wagg\inst{10}
        \and P. P. van der Werf\inst{2}
          }

% institute list. No blank rows allowed
   \institute{Faculty of Electrical Engineering, Mathematics and Computer Science, Delft University of Technology, Delft, The Netherlands\\
         \email{m.rybak@tudelft.nl}
        \and Leiden Observatory, Leiden University, Niels Bohrweg 2, 2333 CA Leiden, the Netherlands
        \and Cosmic Dawn Center (DAWN)
        \and DTU-Space, Technical University of Denmark, Elektrovej 327, 2800 Kgs. Lyngby, Denmark
        \and Department of Physics and Astronomy, University College London,Gower Street, London WC1E 6BT, UK
         \and I.\,Physikalisches Institut, Universit\"at zu K\"oln, Z\"ulpicher Strasse 77, D-50937 K\"oln, Germany
        \and Centro de Astrobiolog\'ia (CAB), CSIC-INTA, Cra. de Ajalvir Km. 4, 28850 Torrej\'on de Ardoz, Madrid, Spain
        \and European Southern Observatory, Karl-Schwarzschild-Stra{\ss}e 2, 85748 Garching bei M\"unchen, Germany
        \and Max Planck Institute for Astronomy, K\"onigstuhl 17, D-69117 Heidelberg, Germany
        \and SKA Observatory, Lower Withington Maccleseld, Cheshire SK11 9FT, UK
        }
   \date{Received 28th April 2022; accepted 19th July 2022}

\abstract{Dusty star-forming galaxies (DSFGs) at redshift $z\geq1$ are among the most vigorously star-forming galaxies in the Universe. However, their dense ($\geq10^5$ cm$^{-3}$) gas phase -- typically traced by HCN(1--0) -- remains almost entirely unexplored: only two DSFGs have been detected in HCN(1--0) to date.
We present the results of a Karl G. Jansky Very Large Array survey of the J=1--0 transition of HCN, HCO$^+$, and HNC(1--0) in six strongly lensed DSFGs at $z=2.5-3.3$, effectively doubling the number of DSFGs with deep observations of these lines.
We detect HCN(1--0) emission in one source (J1202+5354, 4.4$\sigma$), with a tentative HCO$^+$(1--0) detection in another (J1609+6045, 3.3$\sigma$). Spectral stacking yields strict upper limits on the HCN/FIR ($\leq3.6\times10^{-4}$) and HCN/CO(1--0) ratios ($\leq$0.045). The inferred HCN/FIR ratios (a proxy for the star-formation efficiency) are consistent with those in $z\sim0$ far-infrared-luminous starbursts. However, the HCN/CO ratios -- a proxy for the dense-gas fraction -- are a factor of a few lower than suggested by the two previous DSFG detections. Our results imply that most DSFGs have low dense-gas fractions. A comparison with theoretical models of star-forming galaxies indicates that the bulk of gas in DSFGs is at lower densities ($\approx$10$^2$~cm$^{-3}$), similar to 'normal' star-forming galaxies, rather than ultraluminous starbursts.}

    \titlerunning{PRUSSIC I - Dense gas in lensed DSFGs}
    \authorrunning{M. Rybak et al.}
   \maketitle
%
%-------------------------------------------------------------------

\section{Introduction}

Sub-millimetre bright, dusty, star-forming galaxies (DSFGs) with star-formation rates (SFR) of a few $10^2$-$10^3$ $M_\odot$ yr$^{-1}$ play a crucial role at the epoch of the peak star-forming activity of the Universe (redshift $z=$1-4, see \citealt{Casey2014} for a recent review). Though low in numbers, DSFGs account for $\sim$20\% of the total SFR and up to 50\% of the total stellar mass at $z\simeq2$ (e.g. \citealt{Swinbank2014}). Consequently, characterising the star-forming processes in DSFGs is crucial for understanding the stellar mass assembly in the first few billion years after the Big Bang.

Over the last two decades, the interstellar medium (ISM) of DSFGs has been studied extensively in the rest-frame far-infrared (FIR) dust continuum (tracing the star formation) and the CO and [\ion{C}{ii}] emission (tracing the molecular gas), down to sub-kiloparsec scales (see reviews by \citealt{Carilli2013} and \citealt{Hodge2020}). These studies have revealed massive molecular gas reservoirs, often a factor of a few more extended than the FIR-bright starburst (CO: e.g. \citealt{Riechers2011, Dannerbauer2017, Calistro2018}; [\ion{C}{ii}], $z=3-4$: \citealt{Gullberg2018, Rybak2019, Rybak2020b}, $z\geq4$: \citealt{Fujimoto2020, Ginolfi2020}).

Despite these advances in CO and [\ion{C}{ii}] in observations, the actual link between the molecular gas and star formation – the high-density gas – remains largely unexplored. This is because of the low critical density of low-$J$ CO and [\ion{C}{ii}] lines ($n=10^2$ - $10^3$ cm$^{-3}$) which causes them to be collisionally excited across the bulk of the gas reservoir, whereas the stars form in dense cores with $n\geq10^5$ cm$^{-3}$. Conversely, while mid- and high-$J$ CO lines have nominally high critical densities ($n_\mathrm{crit}=1.7\times10^5$ cm$^{-3}$ for CO(5--4)), they can be significantly affected by non-collisional excitation. In fact, the statistical studies of the relation between high-$J$ CO lines and SFR \citep{Greve2014, Liu2015, Kamenetzky2016} show significant discrepancies in the inferred slope, and they differ by $\geq$1~dex in the high-SFR regime, making the connection between high-$J$ CO emission and dense gas difficult to ascertain.

A more direct tracers of dense gas are the emission lines of molecules with high electric dipole, such as HCN, HCO$^+$, and HNC. The ground-state rotational transitions of these molecules have critical densities of $n_\mathrm{crit}\simeq10^5$ cm$^{-3}$ \citep{Shirley2015}. As HCO$^+$ and HNC are more susceptible to excitation by non-thermal processes, HCN(1--0) has emerged as a ``gold standard'' tracer of dense gas \citep{Gao2004b,Bigiel2016,Jimenez2019}. Although Galactic studies show that the HCN(1--0) emission can still occur at densities down to $\sim10^3$~cm$^{-3}$ \citep{Kauffmann2017, Goicoechea2021}, HCN emission is still much better suited for tracing high-density gas than the CO, [\ion{C}{i}] or [\ion{C}{ii}] lines. HCN(1--0) observations provide critical insights into two key questions about star-forming galaxies:
\begin{itemize}
    \item What fraction of the molecular gas is at densities required for star-formation?
    \item How efficiently is the dense gas converted into stars?
\end{itemize}

At $z=0$, HCN(1--0) emission has been extensively surveyed in both main-sequence galaxies and ultra-luminous infrared galaxies (ULIRGs). Following the early single-dish and interferometric studies (e.g. \citealt{Nguyen1992, Solomon1992, Aalto1995}), a ground-breaking survey of more than 50 galaxies by \citet{Gao2004a, Gao2004b} established a linear HCN-SFR correlation in external galaxies\footnote{Yu Gao's seminal work on dense-gas tracers has inspired much of this project. Sadly, Yu Gao passed away during the review of this paper.}. Further observations extended this work to the ULIRG regime (e.g. \citealt{Gracia2008, Krips2008, Garcia2012,Privon2015}), sub-galactic scales (e.g. \citealt{Usero2015, Bigiel2016, Gallagher2018a, Sliwa2017, Jimenez2019}) as well as to adjacent emission lines of HCO$^+$ and HNC (e.g. \citealt{Gracia2006, Costagliola2011}). Together with the observations of individual star-forming clouds in the Milky Way (e.g. \citealt{Wu2005}), these studies have established a linear correlation between HCN and FIR-traced star-formation spanning over 8~dex, from nearby star-forming regions to starburst ULIRGs (see \citealt{Jimenez2019} for a recent summary). This tight ($<0.5$~dex scatter), linear correlation indicates a fundamental relation between the dense gas and star-formation. Conversely, the scatter in the HCN-FIR relation results from varying star-formation efficiency.

How do DSFGs fit into this picture? Unfortunately, detecting HCN(1--0) emission in high-redshift galaxies is challenging due to the intrinsic line faintness (often $\geq$10$\times$ fainter than CO(1--0)), the large luminosity distances, and the need for weather-sensitive cm-wavelength observations. 
Consequently, despite two decades of effort, the HCN(1--0) emission has been detected in only a handful of $z\geq1$ sources: two DSFGs: SMM J16359+6612\footnote{Throughout this paper, we use the updated CO(1--0) luminosity from \citet{Thomson2012}; this decreases the inferred HCN(1--0)/CO(1--0) ratio by $\sim$30\% compared to the \citet{Gao2007} value, from $L'_\mathrm{HCN(1-0)}/L'_\mathrm{CO(1-0)}=$0.15 to 0.11.} \citep[henceforth J16359]{Gao2007} and SDP.9 \citep{Oteo2017}, and three quasar hosts \citep{Solomon2003, vdBout2004, Carilli2005}, all of them gravitationally lensed. In fact, while the number of CO- and [\ion{C}{ii}]-detected DSFGs has been increasing rapidly (c.f.,\citealt{Carilli2013} vs \citealt{Hodge2020}), the number of HCN(1--0) detections remains almost stagnant. Even the number of informative non-detections in DSFGs is very limited: only four DSFGs have published HCN(1--0) upper limits\footnote{SMM J1401+0252 \citep{Carilli2005}, SMM 02396-0134 \citep{Gao2007}, Eyelash \citep{Danielson2013}, and SDP.11 \citep{Oteo2017}.}.

Going up the excitation ladder, the intrinsically brighter mid-$J$/high-$J$ HCN, HCO$^+$ and HNC lines ($J_\mathrm{upp}$=3-5) in $z>2$ DSFGs have been studied at mm-wavelengths, either directly in individual objects (e.g. \citealt{Danielson2013, Oteo2017, Bethermin2018, Canameras2021}) or in spectral stacks \citep{Spilker2014}. However, in contrast to HCN(1--0), mid/high-$J$ emission HCN might be strongly affected by mechanical feedback due to shock dissipation \citep{Kazandjian2012, Kazandjian2015, Papadopoulos2014} or infrared pumping \citep{Aalto2007, Riechers2007, Riechers2010}, which makes their interpretation difficult. The ground-state transitions are thus our best bet at directly probing the dense molecular gas. 

One of the main findings of \citet{Gao2004b} was that in nearby galaxies, the HCN/CO and FIR/HCN ratios increase with FIR luminosity (and SFR, by extension). In other words, intensely star-forming galaxies have higher dense-gas fractions ($f_\mathrm{dense} \propto L'_\mathrm{HCN(1-0)}/L'_\mathrm{CO(1-0)}$) and dense-gas star-formation efficiency (SFE=SFR/$M_\mathrm{gas} \propto L_\mathrm{FIR}/L'_\mathrm{HCN(1-0)}$) than normal galaxies. This conclusion is supported by further studies of HCN and HCO$^+$ emission in ULIRGs. For example, \citet{Gracia2008} and \citet{Garcia2012} found that ULIRGs have HCN/FIR ratio 2-4$\times$ higher than $L_\mathrm{FIR}\leq10^{11}$~$L_\odot$ galaxies, while the HCN/CO ratios increase with $L_\mathrm{FIR}$.

The two extant HCN(1--0) detections in DSFGs seem to conform to this trend, with HCN/CO ratios of 0.11 (J16359) and 0.3 (SDP.9), higher than most ULIRGs. Taken at face value, this would imply that the intense star-formation in DSFGs ``\textit{is associated with more massive dense molecular gas reservoirs and higher dense molecular gas fractions}'' \citep{Oteo2017}. However, given the lack of DSFGs with secure HCN(1--0) detections or informative upper limits, can this statement be extended to DSFGs as a population?

To address this issue, we have initiated the \textsc{Prussic} survey\footnote{``Prussic acid'' is another name for HCN, which was first isolated from the Prussian blue pigment.} - a concerted effort to characterise dense gas tracers in a sizeable sample of high-redshift galaxies using JVLA, ALMA and NOEMA.

In this paper, we present the results of a Karl G. Jansky Very Large Array (JVLA) campaign targeting the $J=1-0$ rotational lines of HCN, HCO$^+$, and HNC in six $z\sim3$ lensed DSFGs, effectively doubling the number of DSFGs with deep HCN, HCO$^+$, and HCN(1--0) observations (from 5 to 11). Our deep JVLA observations allow us to put stringent constraints on the dense-gas fractions and star-formation efficiencies in DSFGs. We focus on the HCN(1--0) line, as this provides the best constraints on the dense gas content of our targets. The upcoming papers based on ALMA and NOEMA data will cover the mid-$J$ HCN, HCO$^+$, and HCN emission lines.

This paper is structured as follows. In Section~\ref{sec:observations}, we outline the JVLA observations and data reduction. Section~\ref{sec:results} discusses the imaging and signal-extraction procedure  and lists the line detections and upper limits. Section~\ref{sec:discussion} discusses the dense gas content and star-formation efficiency of our sample and puts it in the context of both low- and high-redshift observations and theoretical models. 
Throughout this Paper, we use a flat $\Lambda$CDM cosmology from \citet{Planck2016}, and a Chabrier stellar initial mass function (IMF).

\section{Observations}
\label{sec:observations}

\subsection{Target sample}
\label{subsec:sample}

\begin{table*}[t]
\caption{Target list. Individual columns list the source position, source and lens redshift ($z_L$, $z_L$), sky-plane (lensed) FIR and CO(1--0) luminosities, CO(1--0) FWHM and FIR continuum magnification. $L_\mathrm{FIR}$ inferred from modified black-body fits integrated over 8-1000~$\mu$m.}
    \centering
    \begin{tabular}{l|ccccccc}
    \hline
     Source & RA \& DEC & $z_s$ & $z_L$ & $L_\mathrm{FIR}^\mathrm{sky}$ & $L{'}^\mathrm{sky}_\mathrm{CO(1-0)}$ & FWHM CO(1--0)& $\mu_\mathrm{FIR}$\\
     & [J2000] & & & [$L_\odot$] & [K km s$^{-1}$ pc$^{2}$] & [km s$^{-1}$]& \\
     \hline
     SDP.81 & 09:03:11.6 +00:39:07 & 3.042 & 0.23 & $(43\pm0.5)\times10^{12}$ & $(54\pm9)\times10^{10}$ & 435$\pm$54 & 18.2$\pm$1.2\\
     SDP.130 & 09:13:05.4 --00:53:41 & 2.625 & 0.22 & $(32\pm3)\times10^{12}$ &$(22\pm6)\times10^{10}$ & 377$\pm$62 & 8.6$\pm$0.4\\
     HXMM.02 & 02:18:30.7 --05:31:32 & 3.391 & 1.35 &  $(36\pm3)\times10^{12}$ & $(21\pm2)\times10^{10}$ & 540$\pm$40 & 5.3$\pm$0.2 \\
     %\hline
    J0209 & 02:09:41.3  +00:15:59 & 2.553 & 0.20 & $(133\pm40)\times10^{12}$  & $(54\pm19)\times10^{10}$ & 409$\pm$16 & 14.7$\pm$0.3 \\
   J1202 & 12:02:07.6  +53:34:39 & 2.442 & 0.21 &  $(81\pm40)\times10^{12}$ & $(142\pm50)\times10^{10}$ & 602$\pm$21 & 25$^\dagger$\\ 
    J1609 & 16:09:17.8  +60:45:20 & 3.256 & 0.45 & $(156\pm70)\times10^{12}$ & $(334\pm117)\times10^{10}$ & 705$\pm$31& 15.4$\pm$1.0  \\
     \hline
     \multicolumn{8}{l}{    $\dagger$ - derived using a Tully-Fisher argument rather than lens modelling \citep{Harrington2021}.}
    \end{tabular}
    \\
    \justify
    \textbf{Sources:} Redshifts adopted from \citet[SDP.81]{alma2015}, \citet[SDP.130, HXMM.02]{Bussmann2013}, \citet{Geach2015} and \citet[J0209]{Harrington2016}, \citet{Harrington2016} and \citet[J1202 and J1609]{Canameras2018}. $L_\mathrm{FIR}$ (8-1100~$\mu$m) adopted from \citet[SDP.81]{Rybak2020b}, \citet[SDP.130]{Bussmann2013}, \citet[HXMM.02]{Wardlow2013}, and \citet{Harrington2021}. CO(1--0) line luminosities and FWHM: \citet[SDP.81]{Valtchanov2011}, \citet[SDP.130]{Frayer2011}, D.\,Riechers, in prep. (HXMM.02, CO(1--0)), \citet[HXMM.02. CO(3--2)]{Iono2012}, Planck sources - \citet{Harrington2016, Harrington2018, Harrington2021}. Lensing magnifications: \citet[SDP.81]{Rybak2020b}, \citet[SDP.130]{Falgarone2017}, \citet[HXMM.02]{Bussmann2015}, \citet[J0209]{Geach2018}, \citet[J1609]{Canameras2018}. \\

    \label{tab:sources}
\end{table*}

We target six $z=2.4-3.4$ strongly lensed DSFGs identified in the wide-field \textit{Herschel} and \textit{Planck} imaging, for which the HCN(1--0) line falls into the JVLA K-band. Thanks to the lensing magnification ($\mu_\mathrm{FIR}=5-18$), the on-source time required to achieve the target sensitivity is reduced by about two orders of magnitude. 

The three sources selected from the \textit{Herschel} surveys H-ATLAS survey \citep{Negrello2010, Negrello2017} and HerMES \citep{Oliver2012, Wardlow2012} are SDP.81, SDP.130, and HXMM.02 (also known as Orochi, \citealt{Ikarashi2011}). All three targets have HCN(1--0) line observable in the JVLA K-band, robust lens models based on high-resolution ALMA imaging, and existing CO(1--0) measurements.

The three sources selected from the \textit{Planck} imaging are J0209, J1202 and J1609. J0209 was first identified in the Subaru VISTA-CFHT Stripe 82 survey by the \textsc{SpaceWarps} citizen-science project \citep[9io9]{Geach2015}. The remaining two sources are drawn from the \textit{Planck}'s Dusty GEMs -  Gravitationally Enhanced sub-Millimetre Sources (\citealt{Canameras2015}, where J1202 is listed as PLCK\_G138.6+62.0 and J1609 as PLCK\_G092.5+42.9). All \textit{Planck} DSFGs have very high apparent (lensed) FIR luminosity ($L_\mathrm{FIR}\geq10^{14}$~$L_\odot$). Although the observations towards these sources are relatively shallow (1-2 hours per source), they still achieve high sensitivity in HCN/FIR and HCN/CO ratios. 

The positions and properties of individual targets are listed in Tab.~\ref{tab:sources}. All sources except J1202 have robust lens models derived from high-resolution mm-wave continuum imaging; for J1202, the magnification is derived from a Tully-Fisher relation and can carry significant uncertainty \citep{Harrington2021}.

\subsection{VLA K-band observations, data reduction, and imaging}
\label{sec:VLA_observations}

\begin{table*}[t]
\caption{Summary of JVLA observations: dates of observations, on-source time, synthesised beam FWHM, position angle and continuum rms (for naturally weighted images, see Fig.~\ref{fig:vla_cont}).}
    \centering
    \begin{tabular}{l|cccc}
    \hline
     Source & Dates observed & $t_\mathrm{on}$ & Beam FWHM (PA) & $\sigma_\mathrm{cont}$ \\
     & & [hr] & [arcsec (deg)] & [$\mu$Jy/beam] \\
     \hline
     SDP.81 & 2019 Dec 30, 2020 Jan 5 & 4:25   & 4.0$\times$2.9 (87) & 2.0\\
     SDP.130 & 2020 Jan 4, Jan 12, Jan 16 & 7:01 & 4.3$\times$3.1 (84) & 1.5 \\
     HXMM.02 & 2019 Nov 19, Nov 25, Dec 6, Dec 10 & 13:38 & 4.4$\times$3.0 (95) & 2.7$^\dagger$\\
     J0209 & 2017 Feb 16, Feb 22, May 6  & 0:15 & 3.5$\times$2.8 (118) & 12.5\\
     J1202 & 2017 Feb 18, Feb 22 & 0:30 & 3.0$\times$2.0 (2) & 6.9\\
     J1609 & 2017 Feb 18, Mar 5, Mar 7 & 1:23 & 3.6$\times$3.1 (85) & 5.1\\

     \hline
     \multicolumn{5}{l}{$^\dagger$ - Upper sideband only. The lower sideband suffered from a very strong radio interference.}
    \end{tabular}
    \label{tab:vla_obs}
\end{table*}

We have conducted observations with the JVLA in New Mexico, USA as part of the proposals VLA/19B-265 (PI: Rybak, 2019 November to 2020 January) and VLA/17A-362 (PI: Greve, 2017 February to May).

All targets were observed in the K-band using the most compact D-array, consisting of 25 to 27 25-metre antennas. The baselines ranged from 40 to 1000~m, providing sensitivity to structures on 2.8 - 70~arcsec scales at 22~GHz. To observe the HCN, HCO$^+$, and HNC line in a single tuning, we conducted the observations with 3-bit receivers which give a total bandwidth of 8~GHz. The spectral resolution was 2~MHz.

We reduce the JVLA data using \textsc{Casa} version 5.6 \citep{McMullin2007}, using the parallel \texttt{mpicasa} mode. We calibrate each SB separately using the standard JVLA pipeline before concatenating the data and subtracting a constant continuum signal in the $u,v$-plane. Table~\ref{tab:vla_obs} lists the details of the observations and the final beam sizes and rms noise values.

The data quality was very good with the exception of HXMM.02. For this source, three out of four observing blocks have strong RFI in the lower sideband (which contains HCN, HCO$^+$ and HNC). We mitigated the RFI using \textsc{Casa}'s \texttt{tfcrop} flagging mode, followed by manual flagging in the time domain. With 13:30 hr of observing time, this is the deepest observation in our sample; unfortunately, the resulting sensitivity is much worse than expected.

We produce synthesised images using the \textsc{tclean} task, applying the natural weighting, using the H\"ogbom deconvolution with \texttt{fastnoise=False}. The $\sigma_\mathrm{rms}$ of the synthesised images generally agrees with the expected JVLA performance within the flux calibration uncertainty. For the continuum images, we frequency-average all channels that are not affected by atmospheric lines and RFI. For HXMM.02, only the upper sideband is used for the continuum imaging.

\section{Results}
\label{sec:results}

\subsection{VLA K-band imaging}
\label{sec:data_processing}

\begin{figure*}[t]
\begin{centering}
\includegraphics[width=1.0\textwidth, clip=true]{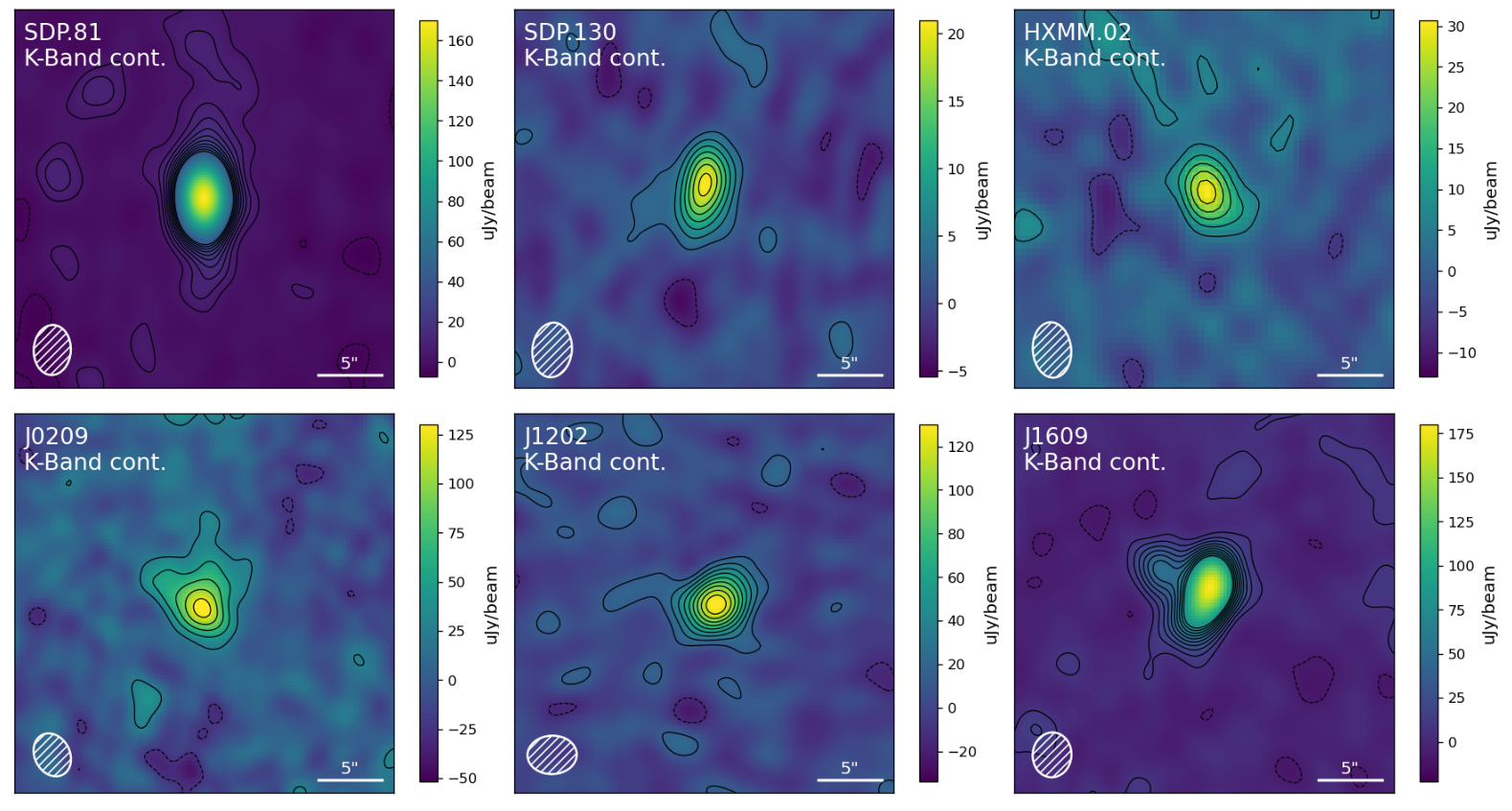}
\\

\caption{VLA K-band imaging: naturally-weighted \textsc{Clean}ed continuum imaging of individual targets. The contours start at $\pm$2$\sigma$, increase in steps of 2$\sigma$ and are truncated at 20$\sigma$. The continuum emission in SDP.81, SDP.130 and HXMM.02 sources is dominated by the synchrotron emission from the AGN in the lensing galaxies. J0209, J1202, and J1609 are all marginally resolved; J1609 shows an extended Einstein ring, most likely free-free emission. 
}
\label{fig:vla_cont}
\end{centering}
\end{figure*}

\begin{figure*}[h]
\begin{centering}
\includegraphics[width=1.0\textwidth, clip=true]{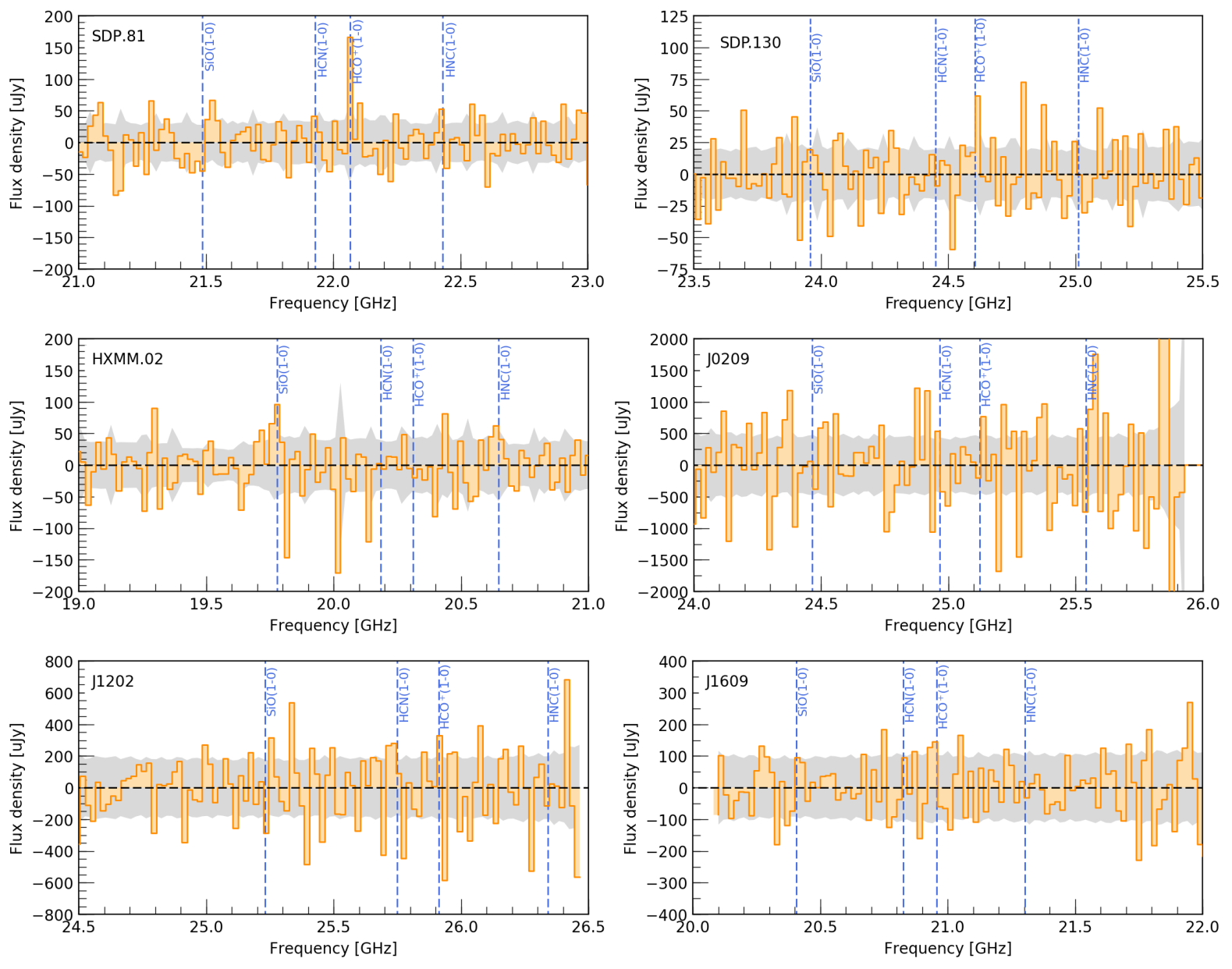}\\

\caption{VLA K-band spectra for individual sources, with the continuum subtracted. The spectrum is extracted from a dirty-image cube (natural weighting) with a channel width of 20~MHz ($\sim$250 km s$^{-1}$). The 1$\sigma_\mathrm{rms}$ level per channel is indicated by the shaded area. No lines are detected at $\geq3\sigma$ significance in individual spectra; in SDP.81, the seemingly strong emission at 22.07~GHz corresponds to a noisier channel (S/N$<$3). For detections from the narrow-band imaging, see Fig.~\ref{fig:vla_narrowband}.}
\label{fig:vla_spectra}
\end{centering}
\end{figure*}

\begin{figure*}[p]
\begin{centering}
\includegraphics[width =0.9\textwidth, clip=true]{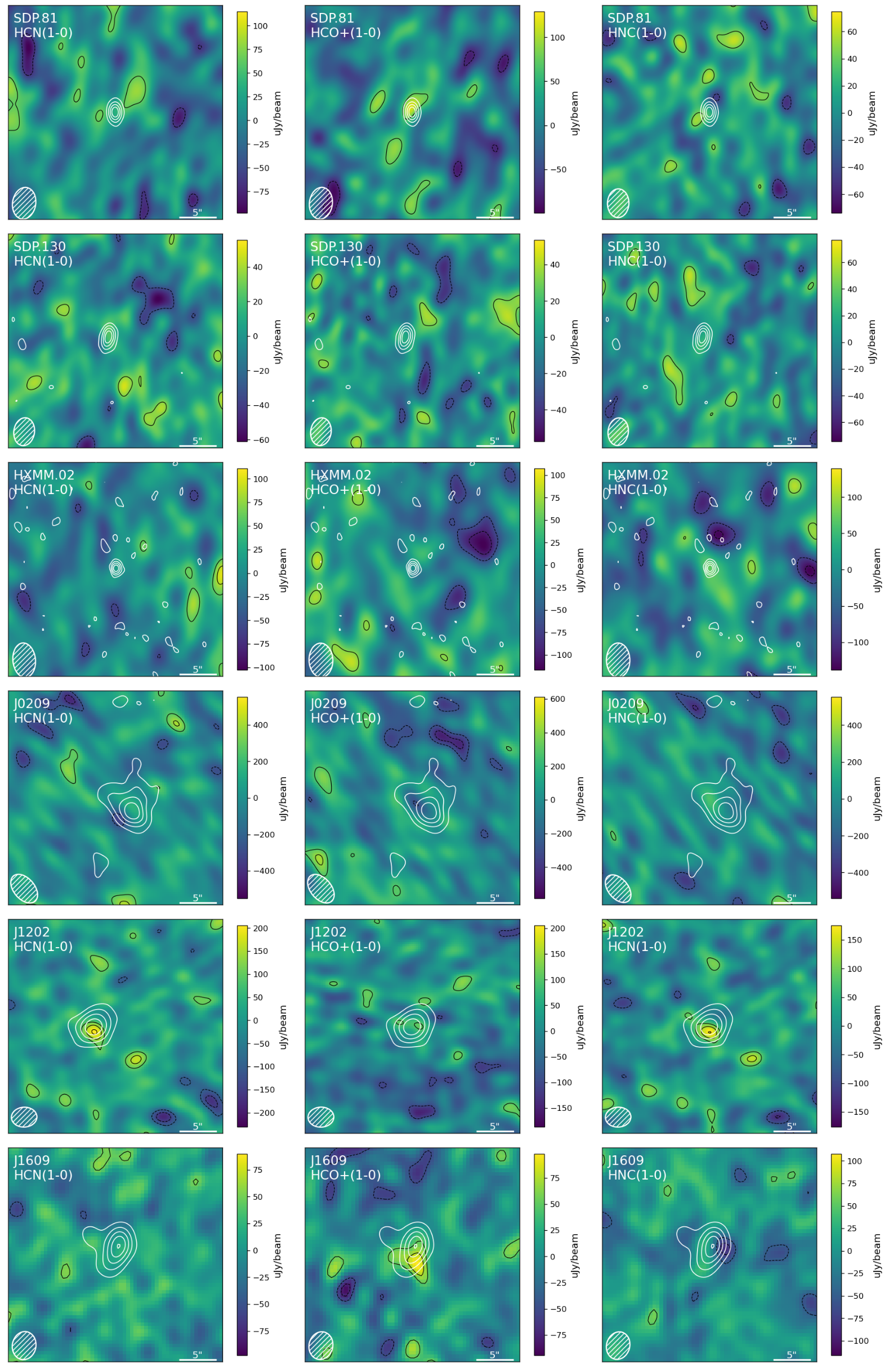}\\

\caption{VLA K-band imaging: narrow-band dirty images, with the K-band continuum indicated by white contours. Black contours start at $\pm2\sigma$ and increase in steps of $1\sigma$.  In J1202, we detect the HCN(1--0) line at 4.6$\sigma$ level; in J1609, we tentatively detect HCO$^+$(1--0) at 3.3$\sigma$. No other lines are detected at $\geq3\sigma$ significance.}
\label{fig:vla_narrowband}
\end{centering}
\end{figure*}

\begin{figure*}
\begin{centering}
\includegraphics[width=18cm, clip=true]{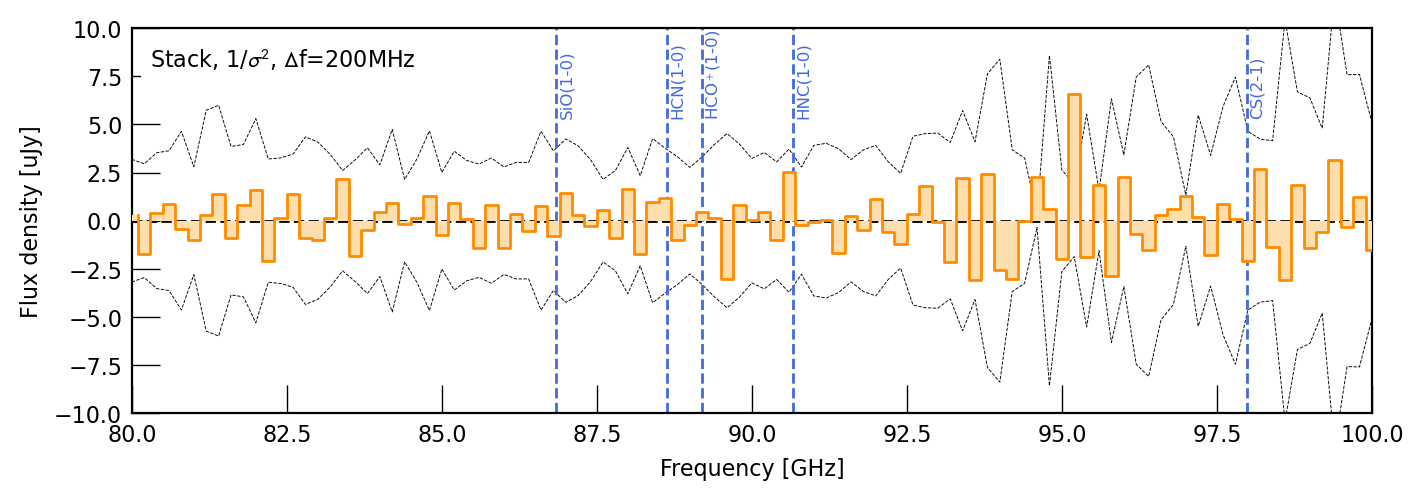}\\
\caption{VLA K-band spectral stack, weighted by $1/\sigma^2$ and re-scaled to $L_\mathrm{FIR}=4.5\times10^{12}$~$L_\odot$ at $z=3$ (mean of our sample, de-lensed). The dashed lines indicate the $\pm3\sigma$ confidence interval. No significant ($\geq3\sigma$) detections are seen in the stacked spectrum.}
\label{fig:vla_stacks}
\end{centering}
\end{figure*}

\subsubsection{K-band continuum}

As shown in Fig.~\ref{fig:vla_cont}, all six targets are significantly detected in the K-band continuum; Tab.~\ref{tab:vla_results} lists the continuum fluxes for individual sources. This emission can be associated either with the background source ($\sim$88~GHz rest-frame) or an AGN in the foreground galaxy. SDP.81, SDP130, and HXMM.02 show continuum emission that is point-like and likely due to the AGN in the lensing galaxy. Namely, the 1.3-cm flux in SDP.81 is consistent with the expected signal from the foreground AGN based on the radio spectrum fit from \citet{Tamura2015}.

On the other hand, the three \textit{Planck}-selected sources show extended continuum emission. J0209 and J1609 are well resolved by the $\sim$3~arcsec beam; their K-band morphology resembles the FIR continuum (see \citealt{Canameras2015}). J1609 in particular shows an extended Einstein arc with a fainter counterimage. In addition, J1202 appears to be marginally resolved. The rest-frame 90~GHz corresponds to the frequency range where the free-free emission is the strongest (compared to the synchrotron and dust continuum); indeed, significant free-free emission has been detected in strongly lensed DSFGs with JVLA \citep{Thomson2012} and ATCA \citep{Aravena2013}.

For the three unresolved sources - SDP.81, SDP.130, and HXMM.02 - we extract the and continuum flux from the \textsc{Cleaned} images; the spectrum is extracted from dirty-image cubes at the position of the peak continuum flux. For the three resolved sources - J0209, J1202, and J1609 - we extract the continuum flux and spectrum using an aperture with 10'' diameter (dictated by the extended continuum structure). 

In addition to our science targets, we detect two serendipitous sources in the K-band continuum. In the SDP.130 field, we detect significant emission (S/N$\geq$10) at J2000 09:13:08.3 -00:53:08.0. This source is robustly detected in the deep VLT/HAWK-I imaging in the ESO archive; no optical emission is seen in the SDSS r, g, b and i filters. In the J1609 field, the first sidelobe of the primary beam contains a very bright point source ($S_\mathrm{1.3cm}$= 787 $\mu$Jy) at J2000 16:09:13.088 60:46:15.956, corresponding to a $z=0.655$ QSO host detected in the JVLA FIRST\footnote{\texttt{http://sundog.stsci.edu/}} imaging at $S_\mathrm{1.4GHz}=9.44\pm0.16$~mJy.

\subsubsection{Emission lines}

We first look for emission lines in the dirty-image cubes. The rest-frame frequencies of individual lines are: HCN(1--0): 88.63~GHz, HCO$^+$(1--0): 89.19~GHz, HNC(1--0): 90.66~GHz. We extract the spectra at the position of the continuum peak for unresolved sources (SDP.81, SDP.130, HXMM.02) or within a 10" diameter aperture for the resolved sources (J0209, J1202, J1609). Figure~\ref{fig:vla_spectra} presents the spectra for individual targets with 20~MHz frequency resolution ($\sim270$ km s$^{-1}$). No lines are detected at $\geq3\sigma$ significance in the individual spectra, independent of the frequency binning used.

Second, we create narrow-band (moment-zero) dirty images by selecting channels within $\pm$FWHM$_\mathrm{CO}$/2 of the systemic frequency of each targeted line\footnote{For the Gaussian line, the area under $\pm$FWHM/2 contains 76\% of the total flux. For the comparison with FIR and CO luminosities in the following section, we scale line fluxes and upper limits reported here by a factor of 1/0.76=1.32.}. The FWHM$_\mathrm{CO}$ is based on the archival CO(1--0) or CO(3--2) linewidths, see Tab.~\ref{tab:sources}. Figure~\ref{fig:vla_narrowband} presents the resulting narrow-band images. We detect significant positive excess for HCN(1--0) emission in J1202 at $S=205\pm49$~$\mu$Jy (4.2$\sigma$) and a tentative HCO$^+$(1--0) detection in J1609 of $83\pm25$~$\mu$Jy (3.3$\sigma$). These point-like detections that be diluted in the 10'' apertures used to extract the spectra in Fig.~\ref{fig:vla_spectra}. No other emission lines are detected. Appendix~\ref{app:A} lists the upper limits on CS(2--1) and SiO(2--1) emission.

\subsection{Spectral stacking}

Besides analysing individual sources, we stack the JVLA spectra extracted from the dirty-image cubes to boost the S/N ratio further. We follow a procedure similar to \citet{Spilker2014}, namely:
\begin{itemize}
    \item Extract spectra for individual galaxies from dirty-image cubes (see Fig.~\ref{fig:vla_spectra}).
    \item Calculate the noise per channel for each dirty-image cube.
    \item Re-scale the observed spectra to a common redshift $z_\mathrm{ref}=3$, i.e. multiplying the observed fluxes by $D_\mathrm{L}(z_\mathrm{source})^2\times(1+z_\mathrm{ref})/(D_\mathrm{L}(z_\mathrm{ref})^2\times(1+z_\mathrm{source}))$.
    \item Combine the individual spectra, using a $1/\sigma(f)^2$ weighting, where $\sigma(f)$ is the rms noise per channel. This maximises the S/N of the stacked spectrum and downweights channels with increased noise due to receiver response or interference.
    \item For each channel, we calculate the weighted-mean $L_\mathrm{FIR}$ and $L'_\mathrm{CO(1-0)}$; we then re-scale the data to a median $L_\mathrm{FIR}=5.3\times10^{12}$~L$_\odot$ of our sample.
\end{itemize}

Figure~\ref{fig:vla_stacks} shows the resulting stack for a rest-frame frequency resolution of 200~MHz ($\sim$670 km s$^{-1}$). No significant emission is detected at the expected line positions. To ensure the robustness of our results, we repeat the analysis using different spectral bin sizes, both for the input data and the stacking. Weighting the spectra by $1/\sigma$ rather than $1/\sigma^2$ %(as in, e.g. \citealt{Boogaard2021}) 
increases the upper limit on HCN/FIR and HCN/CO ratios by a factor of $\sim2$, while HCO$^+$/FIR and HCO$^+$/CO increase by $<$30\%; this increase is driven by the high noise in HXMM.02 and J0209.

We list the inferred line luminosities and rest-frame 88-GHz continuum fluxes in Tab.~\ref{tab:vla_results}. Table~\ref{tab:line_ratios_comparison} lists the corresponding HCN/FIR, HCN/CO, HCO$^+$/FIR, HCO$^+$/HCN, and HNC/HCN ratios for our sample. Although our galaxies are at $z=2.5-3.5$ where the cosmic microwave background temperature increases to $T_\mathrm{CMB}=9.5 - 12.5$~K, this will have only a marginal effect on the observed HCN, HCO$^+$, and HNC line emission and FIR luminosities (see Appendix~\ref{app:B} for a more detailed discussion).

\subsection{Lens models and magnifications}
\label{sec:lens_models}

Due to their low resolution, our JVLA observations can not be directly de-lensed. Instead, we adopt magnification factors based on higher-resolution FIR continuum imaging (see references in Tab.~\ref{tab:sources}). We assume magnifications derived for the FIR continuum, rather than low-/mid-$J$ CO lines or [\ion{C}{ii}]: this is because the low-$J$ CO lines and [\ion{C}{ii}] emission can be significantly more extended than the FIR-bright star-forming source.

Our assumption that the dense gas tracers are co-spatial with the FIR continuum (at least on galaxy-scales relevant here) might introduce systematic bias into the derived line/continuum ratios (e.g. Serjeant, 2012). In particular, if the low-$J$ CO emission is significantly more extended - and less magnified - than the dense gas tracers, the intrinsic global HCN/CO ratios will be overestimated, while the HCN/FIR ratios will be unaffected. This is likely the case in SDP.81, which has a very extended (over $\sim$10 kpc) reservoir of CO(1--0) and [\ion{C}{ii}] gas \citep{Valtchanov2011, Rybak2020b}, whereas the FIR-bright starburst is only 2-3~kpc across and highly magnified.

On the other hand, if the dense-gas tracers follow CO(1--0) rather than the FIR continuum, the intrinsic HCN/CO ratios will be largely unaffected, but the intrinsic HCN/FIR will increase. This might be the case if a significant fraction of the HCN(1--0) emission arises from low-density gas, as seen in some Galactic regions, \citealt{Kauffmann2017}.

To summarise, in the worst-case differential magnification scenarios, the HCN/CO ratios (and the dense-gas fraction) are overestimated, while the HCN/FIR and the star-formation efficiency are under- and over-estimated, respectively.

\begin{table*}[h]
\caption{VLA observations: sky-plane HCN and HCO$^+$ luminosities, and the rest-frame 88~GHz continuum. Upper limits are set at $3\sigma$ level. The HCN(1--0) detection in J1202 and the tentative HCO$^+$(1--0) detection in J1609 are marked in bold; the quoted uncertainties are for the $D=$10\'' circular aperture. Upper limits on additional emission lines are listed in Appendix~\ref{app:A}.}
    \centering
    \begin{tabular}{l|ccccccccc}
    \hline
     Galaxy &
     $L^{\mathrm{sky}}_\mathrm{HCN(1-0)}$ &  $L'^{\mathrm{sky}}_\mathrm{HCN(1-0)}$ &      $L^{\mathrm{sky}}_\mathrm{HCO^+(1-0)}$ &  $L'^{\mathrm{sky}}_\mathrm{HCO^+(1-0)}$   &  $L^{\mathrm{sky}}_\mathrm{HNC(1-0)}$  &  $L'^{\mathrm{sky}}_\mathrm{HNC(1-0)}$& $S_\mathrm{88GHz}$\\
     & [$L_\odot$] & [K km s$^{-1}$ pc$^2$]      & [$L_\odot$] & [K km s$^{-1}$ pc$^2$]  & [$L_\odot$] & [K km s$^{-1}$ pc$^2$] & [$\mu$Jy]\\
     \hline
     SDP.81 & $\leq8.2\times10^5$ & $\leq3.7\times10^{10}$ & $\leq7.2\times10^5$ & $\leq4.2\times10^{10}$ & $\leq4.3\times10^5$ & $\leq2.4\times10^{10}$ & 170$\pm$2 \\
     SDP.130 & $\leq3.1\times10^5$ & $\leq1.4\times10^{10}$ & $\leq2.2\times10^5$ & $\leq1.3\times10^{10}$ & $\leq3.0\times10^5$ & $\leq1.6\times10^{10}$ & 21$\pm$1.4 \\
     HXMM.02 & $\leq8.6\times10^5$ & $\leq5.1\times10^{10}$ & $\leq9.8\times10^5$ & $\leq5.7\times10^{10}$  & $\leq15\times10^5$ & $\leq8.3\times10^{10}$ & 31$\pm$3  \\
     J0209 & $\leq23\times10^5$ & $\leq10\times10^{10}$ & $\leq29\times10^5$ & $\leq13\times10^{10}$ & $\leq19\times10^5$ & $\leq1.1\times10^{10}$ & 266$\pm$28 \\
     J1202 & $\boldsymbol{(18\pm9)\times10^5}$& $\boldsymbol{(8\pm4)\times10^{10}}$ & $\leq9.5\times10^5$ & $\leq4.2\times10^{10}$ & $\leq10\times10^5$  & $\leq5.7\times10^{10}$ & 177$\pm$16\\ 
     J1609 & $\leq10\times10^5$ & $\leq11.4\times10^{10}$ & $\boldsymbol{(12\pm7)\times10^5}$ & $\boldsymbol{(7\pm4)\times10^{10}}$ & $\leq20\times10^5$ & $\leq5.8\times10^{10}$ & 293$\pm$10 \\
     \hline
    \end{tabular}
    \label{tab:vla_results}
\end{table*}

\begin{table*}
\centering
\caption{Ratios of HCN(1--0), HCO$^+$(1--0) and HCN(1--0) (in K km s$^{-1}$ pc$^2$ units) and FIR luminosity (in $L_\odot$) for individual galaxies in our sample and for the stacked spectra (with $1/\sigma^2$ and $1/\sigma$ weighting). Throughout this paper, we adopt upper limits derived using the $1/sigma^2$ weighting.} 
\label{tab:line_ratios_comparison}
    
    \begin{tabular}{lcccccc}
        \hline
        Source & HCN/FIR & HCN/CO & HCO$^+$/FIR & HNC/FIR & HCO$^+$/HCN & HNC/HCN \\%& 
         \hline 
         SDP.81 & $\leq8.5\times10^{-4}$ & $\leq$0.068 & $\leq9.6\times10^{-4}$ & $\leq4.8\times10^{-4}$ & --- & ---  \\
         SDP.130 & $\leq3.3\times10^{-4}$ & $\leq$0.062 & $\leq4.0\times10^{-4}$ & $\leq4.4\times10^{-4}$ & --- & ---  \\
         HXMM.02 & $\leq14\times10^{-4}$ & $\leq0.24$ & $\leq84\times10^{-4}$ & $\leq20.2\times10^{-4}$ & --- & ---  \\
         J0209 & $\leq16.8\times10^{-4}$ & $\leq0.41$ & $\leq17.8\times10^{-4}$ & $\leq13.8\times10^{-4}$ & --- & ---  \\
         J1202 & $(9.8\pm4.9)\times10^{-4}$ & 0.056$\pm$0.028 & $\leq7.2\times10^{-4}$ & $\leq14.4\times10^{-4}$ & $\leq1.7$ & $\leq1.6$ \\
         J1609 & $\leq7.3\times10^{-4}$ & $\leq$0.034 & $(7.5\pm4.3)\times10^{-4}$ & $\leq6.3\times10^{-4}$ & $\geq1.0$ & --- \\
         Stack ($1/\sigma^2$) & $\leq3.6\times10^{-4}$ & $\leq$0.045 & $\leq3.4\times10^{-4}$ & $\leq3.8\times10^{-4}$ & --- & ---\\
         \hline

    \end{tabular}
\end{table*}

\section{Discussion}
\label{sec:discussion}

As shown in Section~\ref{sec:results}, we have obtained only one $>4\sigma$ HCN(1--0) detection across our six targets, alongside a tentative HCO$^+$(1--0) detection. Nevertheless, thanks to the depth of our JVLA data and large magnifications, our non-detections rank among the best constraints on the ground-state HCN, HCO$^+$, and HNC emission in DSFGs.  

We now explore how our results relate to the physics of star formation and previous studies of DSFGs and other high-redshift galaxies\footnote{We adopt the 8--1000~$\mu$m wavelength range for $L_\mathrm{FIR}$. For the bulk of $z\sim0$ and high-z sources, these are derived using a single-temperature modified black-body profile. Notable exceptions are the Cloverleaf quasar whose SED is described by a two-temperature model (e.g. \citealt{Stacey2018}). \citet{Gracia2008} and \citet{Garcia2012} use a two-temperature model, but estimate $L_\mathrm{FIR}$ using only far-IR ($\geq$40~$mu$m) photometry.}.
As we show below, the main implication of the low HCN(1--0) luminosities of our sample is that a significant fraction of DSFGs (potentially the majority) have low dense-gas fractions and a normal- to elevated dense-gas star-formation efficiency: a direct contradiction to previous results based on only two detections \citep{Gao2007, Oteo2017}. In this sense, DSFGs are more akin to $z\sim0$ ``normal'', extended star-forming galaxies, rather than compact ULIRG starbursts.

\subsection{HCN(1--0), HCO$^+$(1--0) and HNC(1--0) versus FIR luminosity}

\begin{figure}
\begin{centering}
\includegraphics[width=0.49\textwidth, clip=true]{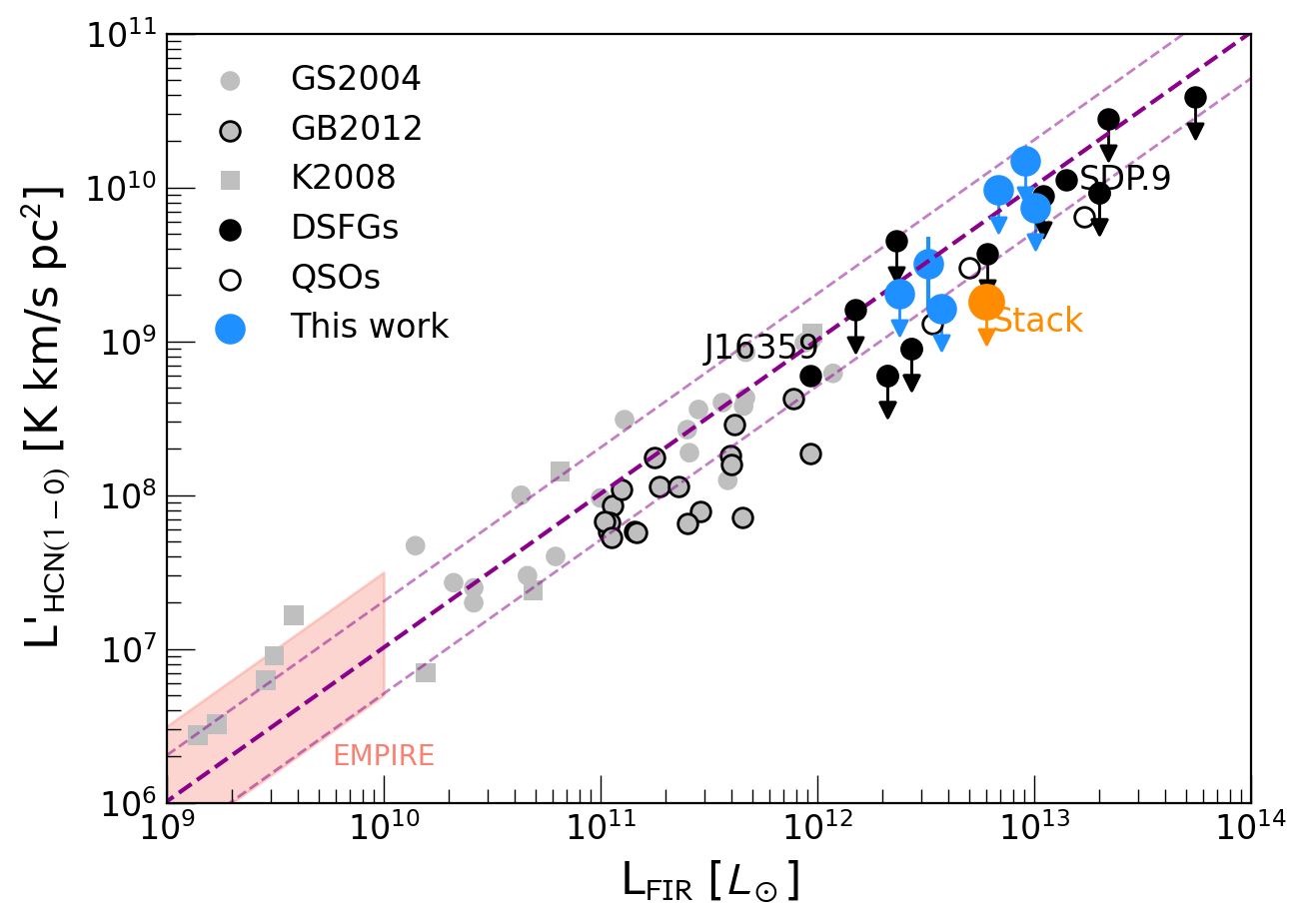}

\caption{Comparison of HCN(1--0) and FIR luminosity in our sample (blue) with low-redshift galaxies \citep{Gao2004a, Krips2008, Garcia2012} and high-redshift DSFGs and QSOs. For the lensed sources, we show the source-plane (de-lensed) values. The upper limit from the stacked spectrum (Fig.~\ref{fig:vla_stacks}) is shown in orange. The purple line indicates the mean HCN/FIR ratio from \citet{Jimenez2019} $\pm$1$\sigma$ scatter (Eq.~\ref{eq:jd19}). The bulk of our sample falls below the linear HCN-FIR trend, in agreement with the expectations of a sub-linear correlation in highly star-forming galaxies \citet{Krumholz2007,Narayanan2008}.}
\label{fig:FIR_HCN_corr}
\end{centering}
\end{figure}

\begin{figure*}
\begin{centering}
\includegraphics[width=0.49\textwidth, clip=true]{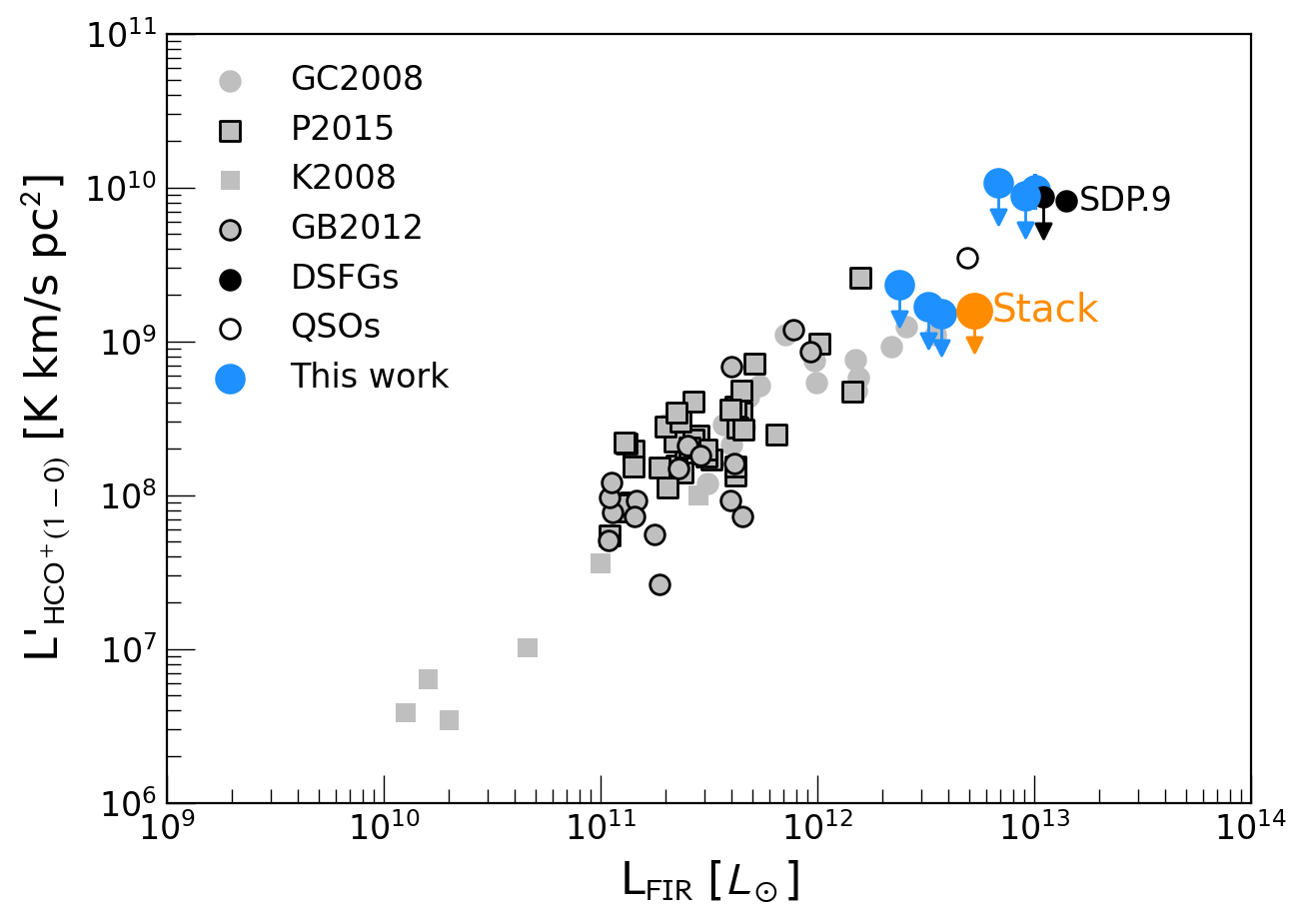}
\includegraphics[width=0.49\textwidth, clip=true]{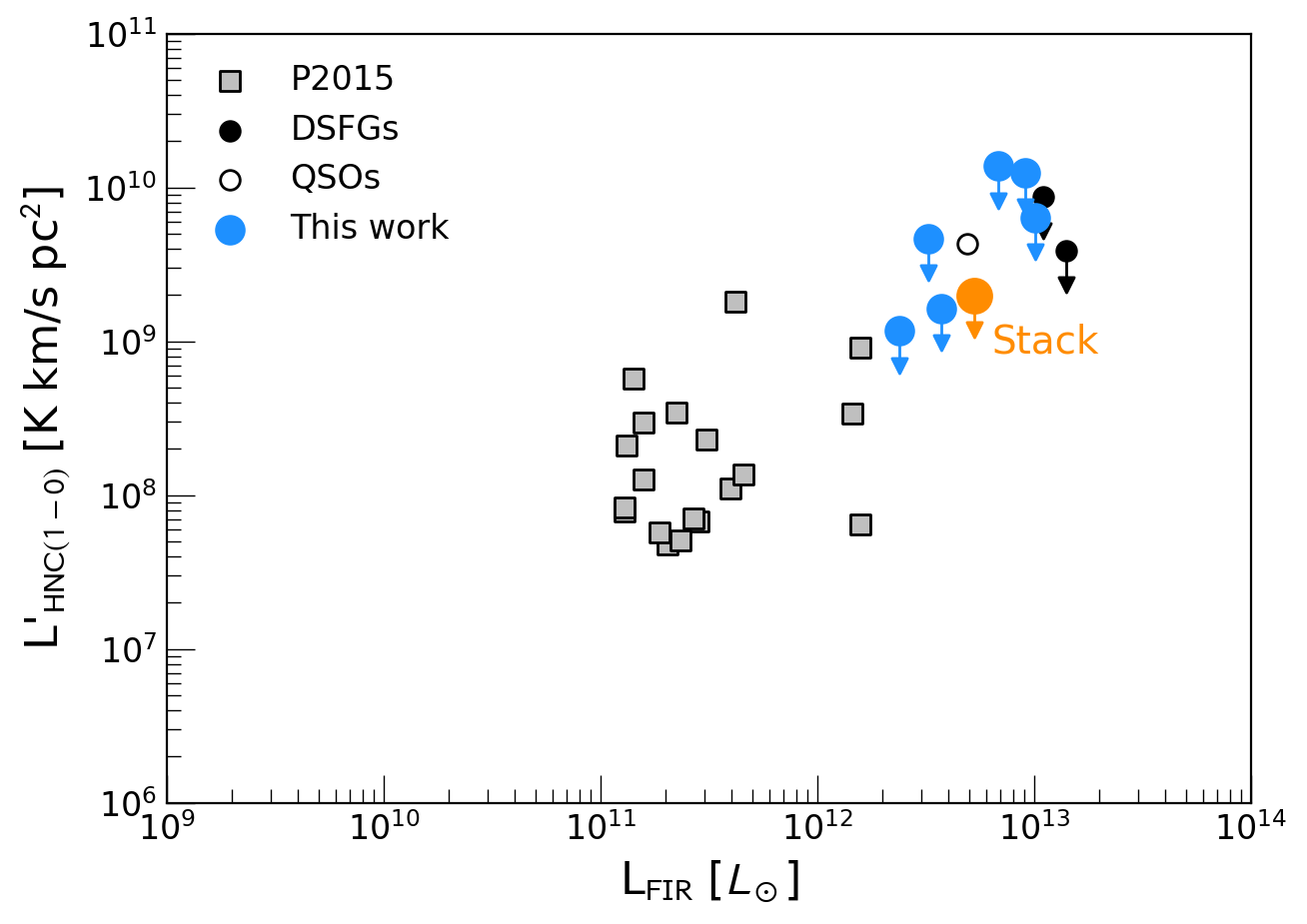}

\caption{As Figure~\ref{fig:FIR_HCN_corr}, but for HCO$^+$(1--0) and HNC(1--0). Notice the increased scatter compared to the FIR-HCN relation and the paucity of observations at high redshift.}
\label{fig:FIR_HCN_corr_2}
\end{centering}
\end{figure*}

Figures~\ref{fig:FIR_HCN_corr} and \ref{fig:FIR_HCN_corr_2} show our HCN, HCO$^+$ and HNC observations in the context of other high-z observations and $z\sim0$ surveys \citep{Gao2004a, Krips2008, Garcia2012, Privon2015}, as well as resolved observations from the EMPIRE survey \citep{Jimenez2019}. As outlined in the introduction, about a dozen high-z galaxies have extant HCN(1--0) observations (mostly upper limits). For the HCO$^+$(1--0) and HNC(1--0) lines, only four high-z galaxies have published detections or upper limits: the Cloverleaf quasar \citep[an HCO$^+$(1--0) detection]{Riechers2006}, the Cosmic Eyelash \citep[non-detections]{Danielson2013}, and SDP.9 and SDP.11 \citep[an HCO$^+$(1--0) detected in SDP.9]{Oteo2017}.

For the HCN(1--0) line, we show the linear $L'_\mathrm{HCN(1--0)}$-$L_\mathrm{FIR}$ trend, obtained by \citet{Jimenez2019} by fitting an extensive compilations of $z=0$ galaxy-integrated observartions:

\begin{equation}
L'_\mathrm{HCN} \, \mathrm{[K\, km s^{-1} \, pc^2]} = \frac{1}{977} L_\mathrm{FIR}\, [L_\odot],
\label{eq:jd19}
\end{equation}

with a 1$\sigma$ scatter of 0.30~dex (\citealt{Jimenez2019}, Tab.~5, for entire galaxies).

The HCN(1--0) detection in J1202 and the upper limits for J0209, SDP.81, and HXMM.02 are within $\pm$1$\sigma$ of the \citet{Jimenez2019} trend (Eq.~\ref{eq:jd19}). However, the HCN/FIR upper limits in SDP.130, J1609, and the stacked spectrum are almost 3$\times$ lower, indicating that HCN/FIR decreases with increasing $L_\mathrm{FIR}$. This agrees with theoretical studies that predict that the $L'_\mathrm{HCN}$-$L_\mathrm{FIR}$ relation becomes sublinear in high gas surface density environments such as galactic centres and DSFGs \citep{Krumholz2007,Narayanan2008}. At the same time, the three individually detected DSFGs - J16359, SDP.9, and J1202 - are consistent with the \citet{Gao2004b} trend. Consequently, while the HCN/FIR ratio might indeed become sublinear at high $L_\mathrm{FIR}$, the source-to-source scatter remains significant ($\sim$1~dex). We discuss the potential enhancements of HCN(1--0) emission due to local conditions in Sections~\ref{subsubsec:HCN_CO_FUV} to \ref{subsubsec:HCN_CO_AGN}.

Finally, we note that the HCN, HCO$^+$, and HNC ratios can be a useful diagnostic of the dense-gas thermodynamics. For example, the HCN/HNC ratios can be used to distinguish between the photon- and X-ray dominated regions (PDR/XDR; for PDRs, HNC/HCN$\leq1$, \citealt{Meijerink2007}), whereas high HCN/HCO$^+$ ratios might indicate the presence of an AGNs \citep{Kohno2005, Papadopoulos2007}. The lack of secure HCN, HCO$^+$, and HNC(1--0) detections in our sources prevents us from using these diagnostics. The upper limits on the HCO$^+$/HCN and HNC/HCN ratios in J1609 merely point towards the PDR regime (HNC/HCN$\leq1.6$)

\begin{figure}
\begin{centering}
\includegraphics[width=0.5\textwidth, clip=true]{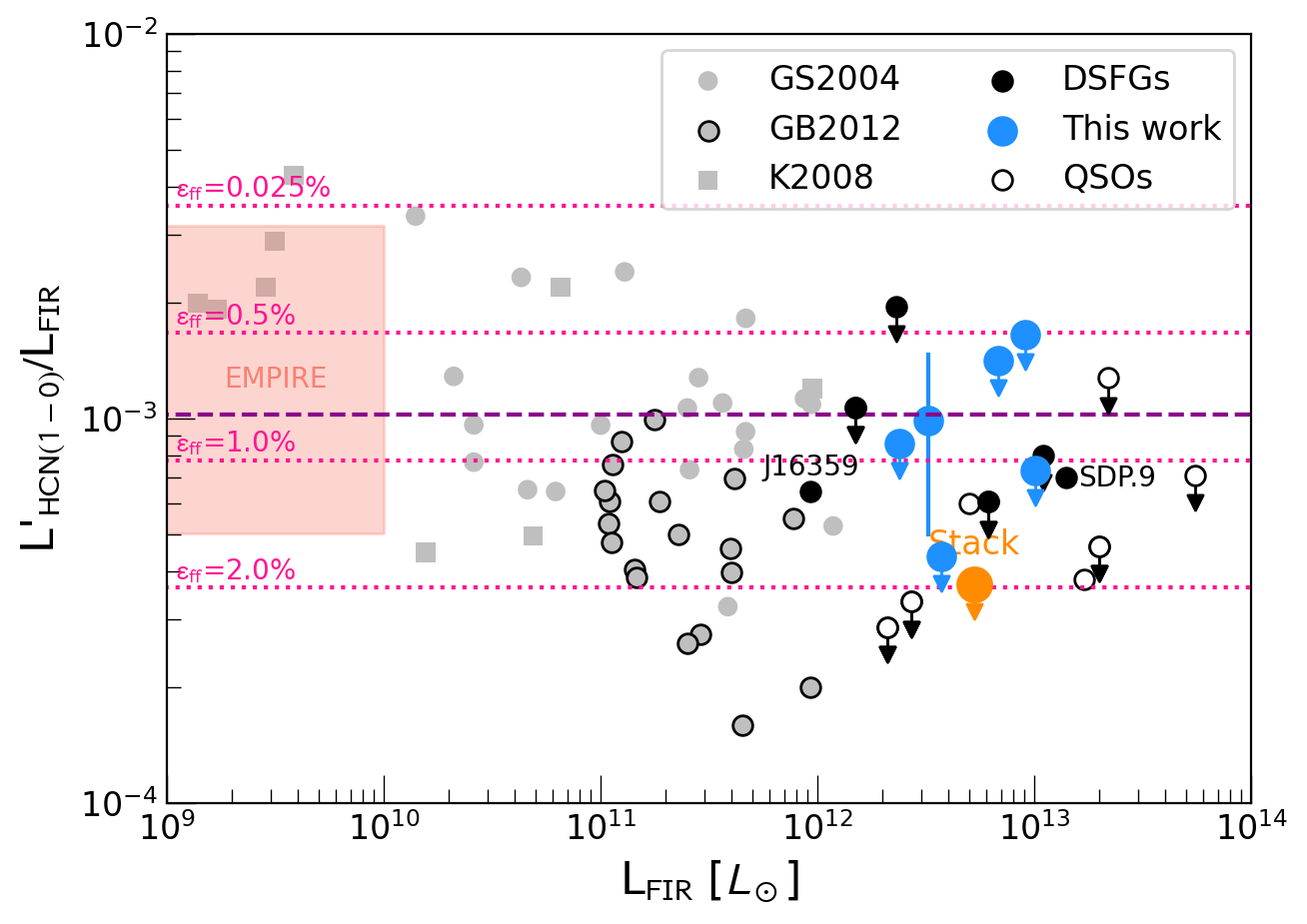}\\
\caption{HCN(1--0)/FIR ratio as a function of FIR luminosity. The pink dashed horizontal lines correspond to star-formation efficiency parameter $\epsilon_\mathrm{ff}$=0.3, 0.5, 1.0. and 2.0\% \citep{Onus2018}; the purple line indicates the HCN/FIR ratio from \citet{Bigiel2016}. We find $\epsilon_\mathrm{ff}=1$\% in J1202, other individual upper limits and the stacked spectra imply $\epsilon_\mathrm{ff}\geq2$\%.}
\label{fig:FIR_HCN}
\end{centering}
\end{figure}

\begin{figure}
\begin{centering}
\includegraphics[width=0.5\textwidth, clip=true]{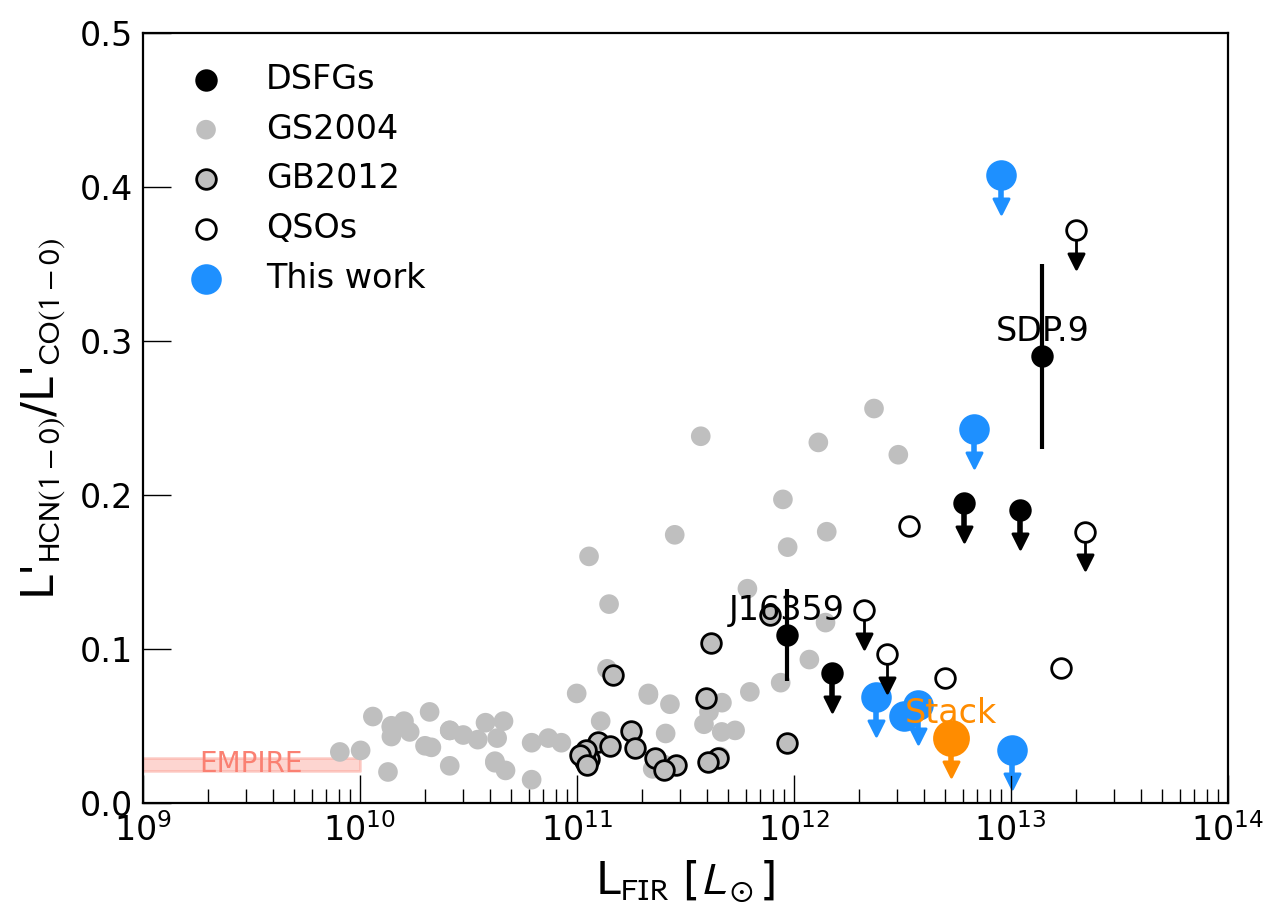}
\caption{Far-infrared luminosity versus dense-gas fraction ($L'_\mathrm{HCN}/L'_\mathrm{CO(1-0)}$) for our sample (blue), compared to other high-redshift DSFGs \citep{Gao2007, Oteo2017} and $z\sim0$ normal star-forming galaxies and ULIRGs from \citet{Gao2004a, Garcia2012}. Our sample shows much lower HCN/CO ratios than reported for SDP.9 and J16359, indicating that most DSFGs have low dense-gas fractions.}
\label{fig:CO_HCN}
\end{centering}
\end{figure}

\subsection{Star-formation efficiency in DSFGs}
\label{subsec:SFE_in_DSFGs}

Our HCN(1--0) measurements can be used to constrain the SFR efficiency per free-fall time $\epsilon_\mathrm{ff}$ (e.g.\,\citealt{KrumholzMcKee2005}):

\begin{equation}
    \epsilon_\mathrm{ff}=\frac{M_\mathrm{gas}}{t_\mathrm{ff}\times \mathrm{SFR}}
    \label{eq:epsilon_ff}
\end{equation}

where $t_\mathrm{ff}$ is the free-fall timescale, $t_\mathrm{ff}=\sqrt{3\pi/(32G\rho)}$, $G$ is the gravitational constant and $\rho$ the mean gas density.

Several studies have used high-resolution simulations on the galaxy and cloud levels to link the HCN/CO and HCN/FIR ratios to $\epsilon_\mathrm{ff}$ (e.g. \citealt{Hopkins2013, Onus2018}). Here, we focus on the cloud-scale simulations of \citet{Onus2018}. Namely, \citet{Onus2018} used hydrodynamic simulations of star-forming clouds with a range of feedback prescriptions to derive an empirical relationship between the SFR/HCN
ratio and $\epsilon_\mathrm{ff}$:
\begin{equation}
\mathrm{SFR}= 2.2\times10^{-7} \left( \frac{\epsilon_\mathrm{ff}}{0.01} \right)^{1.1} L'_\mathrm{HCN(1-0)}.
\end{equation}

Re-writing this in terms of $L_\mathrm{FIR}$, using the SFR-$L_\mathrm{FIR}$ relation (SFR [$M_\odot$ yr$^{-1}$] $=1.71\times10^{-10}\,L_\mathrm{FIR}$ [$L_\odot$], \citealt{Kennicutt1998}) and assuming the Chabrier stellar initial mass function, we obtain:

\begin{equation}
L_\mathrm{FIR}\simeq 1286 \left( \frac{\epsilon_\mathrm{ff}}{0.01} \right)^{1.1} L'_\mathrm{HCN(1-0)}.
\end{equation}

For comparison, the galaxy-averaged value for the Milky Way is $\epsilon_\mathrm{ff}\approx0.5$\% \citep{Murray2011}, whereas $\epsilon_\mathrm{ff}$ reaches up to a few \% in ULIRGs from the \citet{Garcia2012} sample, assuming there is no change in the HCN-dense gas trend. The two HCN (1--0) detections in \citet{Gao2007} and \citet{Oteo2017} imply $\epsilon_\mathrm{ff}=1$~\%. As shown in Fig.~\ref{fig:FIR_HCN}, our upper limits and the detection in J1202 are consistent with $\epsilon_\mathrm{ff}\geq1$\%, with SDP.81 and J1609 implying $\epsilon_\mathrm{ff}\geq2$~\%. Similarly, the upper limit for the stacked spectrum implies $\epsilon_\mathrm{ff}\geq2$~\%, higher than in most $z\sim0$ star-forming galaxies, and several times higher than the Milky Way. Our $\epsilon_\mathrm{ff}$ estimate of a few \% is consistent with the \citet{Krumholz2011} estimates for the \citet{Genzel2010} and \citet{Tacconi2010} samples of high-$z$ starbursts (median (median $\epsilon_\mathrm{ff}$=2.4\%), based on dynamical arguments and an assumption of a volumetric star-forming law. The HCN/FIR ratios in our sample thus provide independent evidence for elevated $\epsilon_\mathrm{ff}$ in DSFGs.

\subsection{Dense gas content of DSFGs}
\label{subsec:DenseGas_in_DSFGs}

As the CO(1--0) emission traces the total molecular gas (down to densities of $n=10^2-10^3$ cm$^{-3}$) whereas HCN(1-0) traces the high-density gas, the HCN(1--0)/CO(1--0) ratio is a useful proxy for the dense-gas fraction $f_\mathrm{dense}=M_\mathrm{dense}/M_\mathrm{gas}$. All our targets have high-quality CO(1--0) observations from either the JVLA (\citealt{Valtchanov2011, Oteo2017}, D. Riechers, priv. comm.) or the Green Bank Telescope \citep{Harrington2021}. This circumvents the potentially significant uncertainty in deriving gas masses from mid-J CO lines.

Figure~\ref{fig:CO_HCN} compares the HCN/CO ratio in our sample to DSFGs from \citet{Gao2007}, \citet{Oteo2017}, and local galaxies. Apart from HXMM.02 (suffering from a strong RFI) and J0209 (short exposure), we find HCN/CO ratios $\leq$0.07. The HCN(1-0) detection in J1202 yields $L'_\mathrm{HCN}/L'_\mathrm{CO}=0.056\pm0.028$, similar to the upper limit for J1609. Finally, our spectral stack implies $L'_\mathrm{HCN}/L'_\mathrm{CO}\leq0.045$ (3$\sigma$ upper limit).

The inferred HCN/CO ratios for our sample are a factor of a few lower than in J16359 ($\approx$0.11) and SDP.9 ($\approx$0.29).
Consequently, the hypothesis that DSFGs have \emph{universally} high molecular gas fractions is not supported by our data. Instead, our results indicate that \emph{most DSFGs have low dense-gas fractions}.

The low HCN(1--0)/CO(1--0) ratios in DSFGs pose a challenge for surveys of dense gas at high redshift. Assuming an HCN/CO ratio of 0.045 and a CO(1--0)-FIR relation from \citet{Kamenetzky2016}, an unlensed $z=3$ DSFG with a star-formation rate of 500~$M_\odot$ yr$^{-1}$ will have an HCN(1--0) flux of 5~$\mu$Jy (assuming line FWHM=500 km s$^{-1}$; for a galaxy with SFR=100~$M_\odot$ yr$^{-1}$, this reduces to $\sim$1~$\mu$Jy. A 5$\sigma$ detection with the JVLA would require $\sim$3000~hr on-source for the former, and $\sim$90000 for the latter. Even with the Next Generation Very Large Array (ngVLA), such observations in unlensed galaxies will remain challenging: a 5$\sigma$ detection would require $\sim$10~hr and 200~hr, respectively (c.f., \citealt{Casey2015, Decarli2018b}. If our results can ve applied to high-z galaxies as a whole, detecting the ground-state HCN, HCO$^+$, and HNC emission in all but the brightest unlensed galaxies will remain beyond the reach of ngVLA.

\subsection{CO and HCN star-formation efficiencies}

\begin{figure*}
\begin{centering}
\includegraphics[height=6.5cm, clip=true]{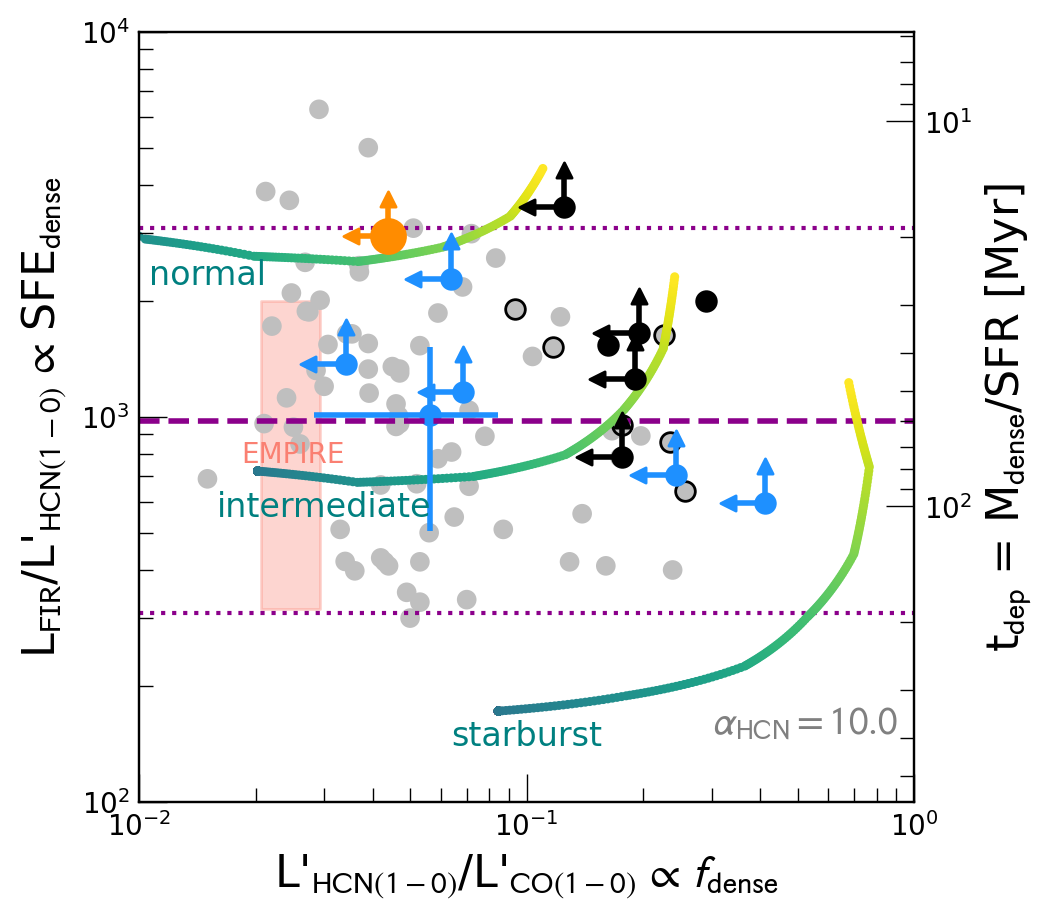}
\hspace{0.5cm}
\includegraphics[height=6.5cm, clip=true]{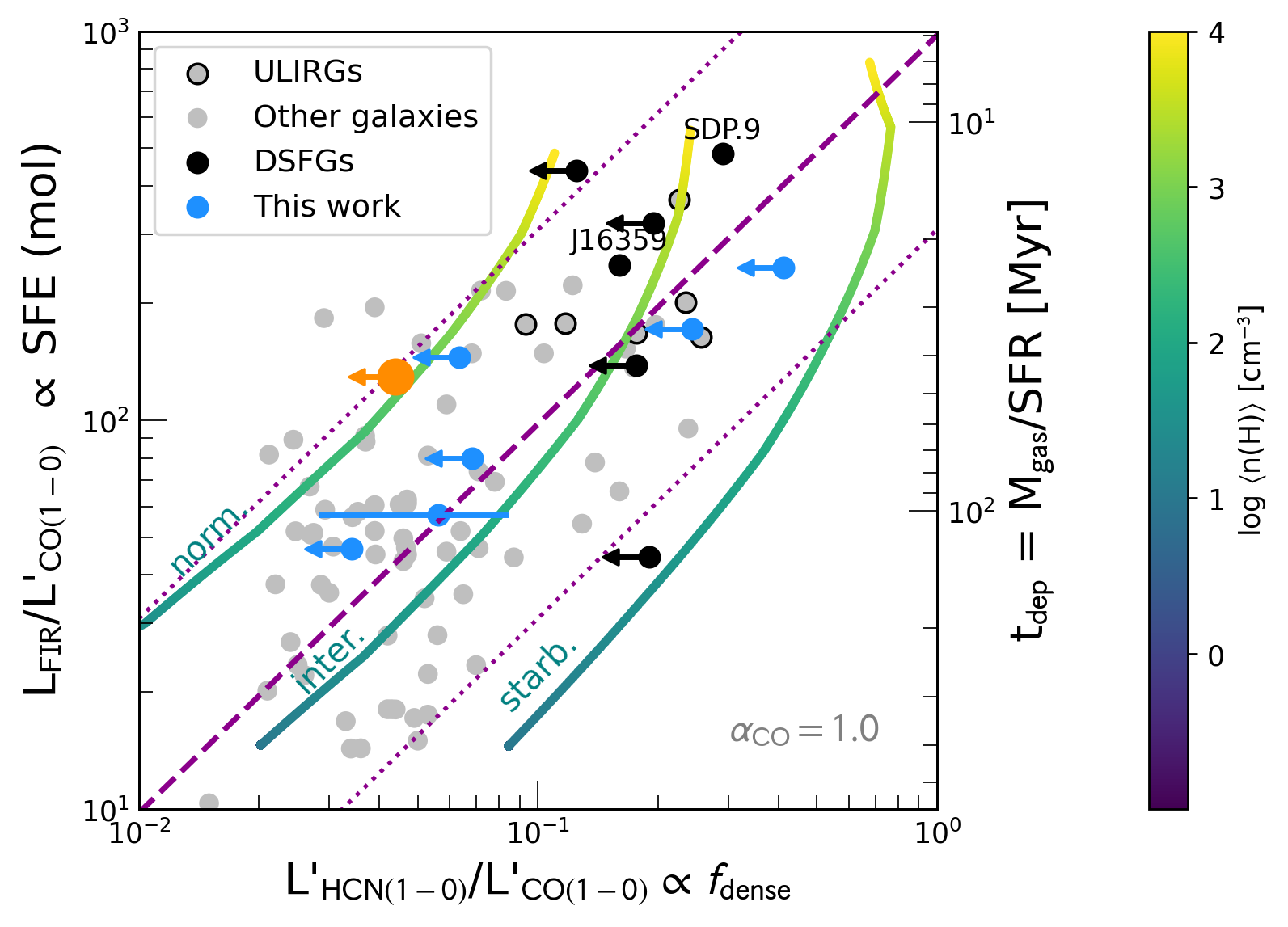}

\caption{FIR/HCN vs HCN/CO ratios (\textit{left}) and FIR/CO vs HCN/CO (\textit{right}) ratios for our sample, other high-$z$ DSFGs and selected $z\sim0$ galaxies \citep{Gao2004a, Garcia2012}; i.e. star-formation efficiency of the molecular / dense gas versus dense-gas fraction (modulo $\alpha_\mathrm{CO}$ and $\alpha_\mathrm{HCN}$). The purple dashed/dotted lines indicate the \citet{Jimenez2019} HCN-FIR relation $\pm$1$\sigma$ scatter. We calculate depletion times $\theta_\mathrm{dep}$ using $\alpha_\mathrm{CO}=1.0$ and $\alpha_\mathrm{HCN}=10.0$ although the actual values vary a factor of a few between different galaxy populations. At a given FIR/CO ratio, DSFGs (and ULIRGs) are systematically offset towards lower HCN/CO. The coloured lines indicate the line ratios for the ``normal'', ``intermediate'', and ``starburst'' models of \citet{Krumholz2007} as a function of mean cloud density $\langle n \rangle$. The DSFG stack is consistent with the ``normal'' model and $\langle n \rangle\leq10^3$ cm$^{-3}$, indicating that the bulk of molecular gas in DSFGs is at low densities and/or with a relatively low Mach number.}
\label{fig:SFE_fdense}
\end{centering}
\end{figure*}

What do our data tell us about the star-forming efficiencies and depletion timescales of the dense and total molecular gas?  Figure~\ref{fig:SFE_fdense} shows the $L_\mathrm{FIR}/L'_\mathrm{HCN(1-0)}$ and $L_\mathrm{FIR}/L'_\mathrm{CO(1-0)}$ (a proxy for molecular/dense-gas star-formation efficiency) versus $L'_\mathrm{HCN(1-0)}/L'_\mathrm{CO(1-0)}$ (a proxy for dense-gas fraction $f_\mathrm{dense}$) for our sample, DSFGs from \citet{Gao2007, Oteo2017}, and $z=0$ galaxies from \citet{Gao2004a, Garcia2012}. For simplicity, we calculate SFE$_\mathrm{dense}$ and SFE$_\mathrm{mol}$ using $\alpha_\mathrm{CO}=1$ and $\alpha_\mathrm{HCN}=10$, although in reality these factors vary by a factor of a few among different galaxy populations.

Two main trends are seen in Fig.~\ref{fig:SFE_fdense}: (1) the dense-gas star-forming efficiency (FIR/HCN) tends to decrease with $f_\mathrm{dense}$; (2) molecular star-forming efficiency (FIR/CO) increases with $f_\mathrm{dense}$, although the scatter is considerable  (as seen in local galaxies, e.g. \citealt{Usero2015, Jimenez2019}). Our three DSFGs with the lowest HCN/CO ratios (SDP.130, J1202, and J1609), as well as our stack, have FIR/CO ratios a factor of 2--7 lower than either J16359 or SDP.9, while having comparable FIR/HCN ratios (Fig.~\ref{fig:FIR_HCN}). In terms of dense-gas $\tau_\mathrm{dep}$, the detection in J1609 implies $\tau_\mathrm{de}^\mathrm{dense}\simeq$60~Myr and comparable to the CO(1-0) depletion timescales ($\sim$100~Myr), but the stacked upper limit pushes $\tau_\mathrm{de}^\mathrm{dense}$ to $\leq20$~Myr, compared to $\sim$80~Myr for the molecular gas.

The combination of low HCN/CO and FIR/CO, but "standard" HCN/FIR ratios seen in DSFGs might be explained if a large fraction of CO(1--0) is not spatially associated with the FIR-traced star-formation. Indeed, resolved studies of dust continuum and molecular gas in DSFGs have found that FIR-bright starburst is often embedded in a much more extended low-$J$ CO (e.g. \citealt{Riechers2011, Calistro2018}) and [\ion{C}{ii}] emission (e.g. \citealt{Gullberg2018, Rybak2019, Rybak2020b}), even in galaxies with narrow linewidth \citep{Frias2022}. For example, SDP.81 has a compact FIR-bright starburst ($\sim$2~kpc across) embedded in a 15-kpc CO(1--0) and [\ion{C}{ii}] reservoir \citep{Rybak2015b, Rybak2020b}. The difference in continuum and low-$J$ CO/[\ion{C}{ii}] sizes might be due to temperature and optical depth gradients \citep{Calistro2018}, accretion of satellites or outflows \citep{Pizzati2020}. 
However, out of the DSFGs considered here, only SDP.81 and J16259 have resolved CO or [\ion{C}{ii}] imaging \citep{Valtchanov2011, Rybak2020b, Thomson2012}, and the line width of CO(1-0) spectra does not show clear evidence for major-mergers in our sample (Fig.~\ref{fig:HCN_CO_tests}). Moreover, as discussed in Section~3.3, very extended cold gas reservoirs would still imply locally low HCN/CO ratios, as the extended CO(1--0) might be on average less magnified than the compact HCN(1--0).

\subsection{Comparison with Krumholz \& Thompson (2007) models}
\label{subsec:KT07}

Constraining the dense-gas fraction and molecular/dense-gas star-formation efficiencies (as seen in Fig~\ref{fig:SFE_fdense}) provides a useful comparison to theoretical models of star-formation.  Broadly speaking, there are two competing models: the density-threshold and the turbulence-regulated models.

The density-threshold model assumes that star-formation rate depends directly on the amount of dense gas available \citep{Lada2010, Lada2012, Evans2014}. In other words, the dense-gas star-forming efficiency (SFE$_\mathrm{dense} \propto L'_\mathrm{HCN}/L_\mathrm{FIR}$) is approximately constant across a wide range of environments, a direct consequence of the linear HCN-FIR relation (e.g. \citealt{Gao2004b,Wu2005}. On the other hand, the total molecular gas star-forming efficiency (SFE$_\mathrm{mol} \propto L'_\mathrm{CO}/L_\mathrm{FIR}$) can vary considerably due to differences in dense-gas fraction ($f_\mathrm{dense} \propto L'_\mathrm{HCN}/L'_\mathrm{CO}$). However, the density-threshold models have been challenged by detailed studies of nearby galaxies (e.g. \citealt{Usero2015, Bigiel2015, Bigiel2016, Jimenez2019}), which show that SFE$_\mathrm{dense}$ depends on the local ISM conditions. In particular, SFE$_\mathrm{dense}$ increases with the stellar surface density $\Sigma_\star$, whereas $f_\mathrm{dense}$ decreases with $\Sigma_\star$.

In contrast, the turbulence-regulated model(s) assume that SFE$_\mathrm{dense}$ depends on the local ISM conditions, particularly the density contrast (peak vs mean gas density), turbulence (forcing mechanism and the Mach number $\mathit{M}$) and, to a lesser extent, the magnetic field \citep{Federrath2012}. Different forms of turbulence-regulated models have been proposed by, for example, \citet{KrumholzMcKee2005, Padoan2011} and \citet{Hennebelle2011}.

We compare our data to the turbulence-regulated model of \citet{Krumholz2005}, which includes the effects of intra-cloud turbulence (but not magnetic fields). A similar comparison has been performed for resolved observations of nearby galaxies by \citet{Usero2015}. Specifically, we consider the HCN(1--0), CO(1--0), and FIR luminosities predicted by \citet{Krumholz2007}, who considered a log-normal gas density distribution for a ``normal'' (Milky-Way like), ``intermediate'', and ``starburst'' galaxy, with increasing gas kinetic temperature ($T=10-50$~K), Mach number ($M=30-80$), metallicity, and optical depth ($\tau_\mathrm{CO(1-0)}=10-20$, $\tau_\mathrm{HCN(1-0)}=0.5-2$). In Fig.~\ref{fig:SFE_fdense}, three models are indicated by viridis-coloured curves.

How do individual DSFGs compare to the Krumholz \& Thompson model? The HCN-bright J16359 and SDP.9 are consistent with the \citet{Krumholz2007} ``intermediate'' model, however, for four DSFGs out of our sample, the ``normal'' model is preferred (HXMM.02 and J0209 have only very weak upper limits). Crucially, for the stacked spectrum, the ``normal'' model with a mean density $\langle n \rangle\leq10^3$ cm$^{-3}$ is strongly preferred. Consequently, this comparison indicates that the star-formation in DSFGs from our sample is similar to that in nearby main-sequence galaxies, rather than starburst ULIRGs.

We note that the agreement with the ``normal'' Krumholz \& Thompson model depends primarily on the HCN/FIR ratio; changing $L'_\mathrm{CO(1-0)}$ would move our data parallel to the model tracks, i.e. to somewhat higher $\langle n \rangle$ values. Finally, we note that the \citet{Krumholz2007} models predict that SFR per free-fall time decreases with increasing Mach number (SFR$_\mathrm{ff}\propto \mathit{M}^{-0.3}$). However, there is mounting evidence that the ISM in DSFGs is highly turbulent (e.g. \citealt{Sharda2018, Dessauges2019, Harrington2021}). Conversely, more recent turbulence-regulated models assume that SFR$_\mathrm{ff}$ increases with turbulence \citep{Federrath2012}. Given the weak dependence of predicted SFE$_\mathrm{dense}$ and SFE$_\mathrm{mol}$ on the Mach number in the Krumholz \& Thompson model, we still consider this comparison relevant.

\subsection{Comparison with HCN(1--0) and HCO$^+$(1--0) detections in high-$z$ quasar hosts}

With the exception of SDP.9, the DSFGs considered in this Paper are dominated by star-formation, with no evidence of significant AGN activity. The presence of an AGN can impact the HCN, HCO$^+$, and HNC excitation in several ways. In general, AGN-dominated systems at $z\sim0$ have elevated HCN/CO and HCN/HCO$^+$ ratios (e.g. \citealt{Krips2008, Privon2015}), potentially due to the X-ray-driven chemistry (although this link might be contentious, see \citealt{Privon2020}); mid-IR pumping of the HCN and HCO$^+$ bending modes; depletion of the CO-traced molecular gas reservoir; or mechanical heating due to supernovae (e.g. \citealt{Costagliola2011, Kazandjian2015}). 

To date, only three $z\geq1$ quasar (QSO) hosts have been detected in the HCN(1--0): the ``Cloverleaf'' (H1413+117; $z=2.5$, \citealt{Solomon2003}), IRAS F10214+4724 ($z=2.3$ \citealt{vdBout2004}), and VCV J1409+5628 ($z\approx2.6$, \citealt{Carilli2005}). All three HCN(1--0)-detected QSOs host substantial star-formation (SFR up to 1000~$M_\odot$ yr$^{-1}$; \citealt{Beelen2004, Stacey2018}). Upper limits for five additional high-$z$ QSO hosts were presented in \citet{Carilli2005, Gao2007}.

Looking at Fig.~\ref{fig:FIR_HCN} and \ref{fig:CO_HCN}, the HCN/FIR and HCN/CO ratios in QSOs fall within the range spanned by the DSFGs, without indications of a substantial boosting of the HCN/FIR ratios by the AGN. The three HCN(1--0)-detected QSOs have relatively high HCN/CO ratios (0.08-0.18), that is, comparable to those in J16359 and SDP.9, rather than DSFGs from our sample. While the elevated HCN/CO ratios in QSOs might be due to non-thermal excitation mechanisms, they might also be explained by their compact sizes: \citet{Stacey2021} have shown that the FIR continuum in $z\sim2$ QSOs (including the Cloverleaf) is more compact and significantly more centrally concentrated than in DSFGs.

\subsection{Comparison with mid-J HCN and HCO$^+$ studies of high-$z$ DSFGs}

In addition to the observations of the ground-state HCN, HCO$^+$, and HNC transitions in DSFGs, several studies have targeted the mid-$J$ rotational lines (i.e. $J_\mathrm{upp}=3,4,5$), either in individual objects \citep{Danielson2013, Oteo2017, Bethermin2018, Canameras2021} or via spectral stacking \citep{Spilker2014}. The mid-$J$ lines are generally brighter than the $J_\mathrm{upp}=1$ line and, at high redshift, fall into the easily accessible 3-mm atmospheric window. However, mid-$J$ lines might be susceptible to mid-IR pumping and mechanical heating by shocks. Moreover, it is still unclear whether the linear correlation between the HCN(1--0) and SFR can be extended to the mid-J HCN lines: different theoretical and observational studies indicate either a linear (e.g. \citealt{Zhang2014, Tan2018}) or sub-linear slope (e.g. \citealt{Narayanan2008, Gracia2008}.

\begin{figure}[h]
\includegraphics[width=9.25cm, clip=true]{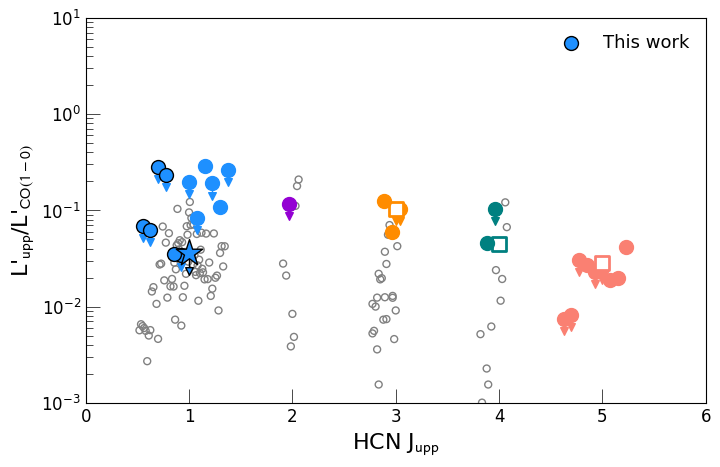} \\
\includegraphics[width=9.25cm, clip=true]{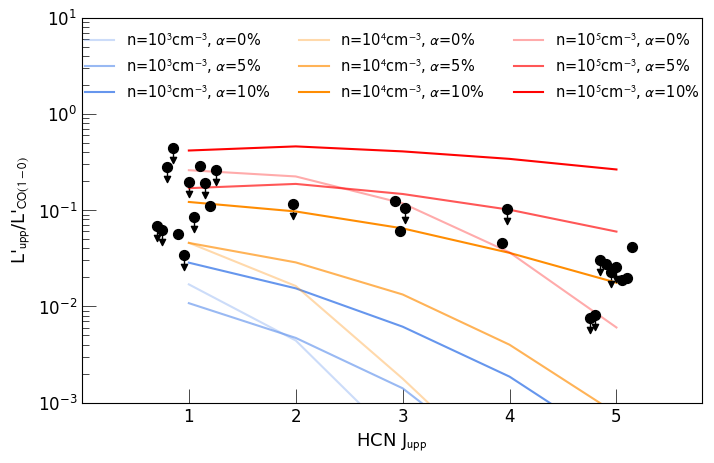}
\caption{Comparison of $L'_\mathrm{HCN}$/$L'_\mathrm{CO(1-0)}$ ratios in DSFGs for HCN(1--0) to (5--4) observations from the literature (\textit{upper}). The coloured points indicate high-z measurements; grey ones $z=0$ observations. Our HCN(1--0) measurements are outlined in black; the star indicates the limit from the stacked K-band spectrum. 
\textit{Lower}: High-z HCN measurements, compared to the predictions of PDR models from \citet{Kazandjian2015}. We vary the PDR gas density ($n=10^3-10^5$~cm$^{-3}$) and the mechanical heating factor ($\alpha$=0-10\%). The FUV irradiation is fixed to $G=10^4$~$G_0$. The current high-z HCN detections can not be reproduced by models with $n\leq10^4$~cm$^{-3}$ without significant mechanical heating.
High-z data: HCN(1--0): \citet{Carilli2005, Gao2007, Oteo2017}; HCN(2--1): G. Jolink (BSc thesis); HCN(3--2): \citet{Danielson2013, Oteo2017}; HCN(4--3) and (5--4): \citet{Bethermin2018, Canameras2021}. Low-z data: HCN(1--0): \citet{Gao2004a, Garcia2012}; HCN(2--1): \citet{Krips2008}, HCN(3--2): \citet{Bussmann2008, Li2020}, HCN(4--3): \citet{Zhang2014}. The squares correspond to the \citet{Spilker2014} stacking analysis.
}
\label{fig:compilation_HCN_FIR}
\end{figure}

Figure~\ref{fig:compilation_HCN_FIR} (upper panel) shows the compilation of existing HCN $J_\mathrm{upp}=1-5$ observations of DSFGs normalised by the CO(1--0) luminosities.
The current high-z HCN(3--2) and (4--3) detections in DSFGs are on the upper end of the $z\sim0$ measurements, both in HCN/FIR and HCN/CO. Given the paucity of high-z measurements, it is unclear if the elevated HCN luminosities in DSFGs are representative, or biased towards the HCN-bright sources.

Out of our sample, only J1609 is detected in the HCO$^+$(5--4) transition with $L'_\mathrm{HCO^+(5-4)}=(5.9\pm0.5)\times10^{10}$ K km s$^{-1}$ pc$^2$ \citep{Canameras2021}. This is comparable to our inferred $L'_\mathrm{HCO^+(1-0)}=(7\pm4)\times10^{10}$ K km s$^{-1}$ pc$^2$. The implied $L'_\mathrm{HCO^+(5-4)}/L'_\mathrm{HCO^+(4-3)}$ ratio of $0.9\pm0.6$ suggests that the HCO$^+$ might be thermalised;  the HCO$^+$ excitation might be higher than even in the Cloverleaf quasar ($L'_\mathrm{HCO^+(4-3)}/L'_\mathrm{HCO^+(1-0)}\approx0.5$, \citealt{Riechers2011b}). 

\citet{Canameras2021} also provide upper limits on the HCN(4--3) and HCO$^+$(4--3) emission in J1202; however, these are substantially higher than our limits for HCN(1--0) and  HCO$^+$(1--0). The remaining four sources in our sample are not sampled in any of the $J_\mathrm{upp}\geq2$ transitions, although ALMA and NOEMA observations of the $J_\mathrm{upp}=3,4$ lines in these sources are currently underway (M. Rybak, in prep.).

For the HCN(5--4) line, there are ten DSFGs with reported detections or informative upper limits \citep{Bethermin2018,Canameras2021}. There are no $z\sim0$ HCN(5--4) data to compare with (the HCN(5--4) line is difficult to observe from the ground, and is outside of, e.g. \textit{Herschel} frequency coverage). Similar to the HCN(1--0) data discussed in this paper, the HCN(5--4) sample shows a $\approx$1~dex scatter in HCN/FIR and HCN/CO(1--0) ratios, indicating a wide range of dense-gas fractions and/or excitation mechanisms. 

Are the HCN observations in DSFGs compatible with a purely collisional excitation? In the lower panel of Fig.~\ref{fig:compilation_HCN_FIR} we compare them against predictions from the photodissociation region models of \citet{Kazandjian2015}. These PDR models assume a simple 1D semi-infinite slab geometry with uniform density. In addition to the photoelectric heating, they specifically incorporate mechanical heating due to external turbulence, which is parametrised by a ratio of photoelectric and mechanical heating at the surface of the cloud, $\alpha$. We consider models with gas density $n=10^3-10^5$~cm$^{-3}$ and $\alpha=0-10$\%. We fix the incident FUV radiation to $G=10^4$~$G_0$; varying $G$ by 1~dex impacts the predicted HCN/CO ratios by a factor of $\sim2$. We assume solar metallicity and the calculation stopping depth of $A_V=10$.

We find that the models with $n\leq10^4$~cm$^{-3}$ can not match the high-J HCN detections without a substantial amount of mechanical heating ($\alpha\geq$10\%). This corroborates a recent claim by \citet{Harrington2021} who find significant mechanical feedback in DSFGs from modelling of the CO ladders. However, if the density in star-forming regions exceeds $10^5$~cm$^{-3}$ (as is the case in SDP.81, c.f., \citealt{Rybak2020b}), the mechanical feedback contribution is reduced to less than a few \%. In other words, reproducing the current high-redshift HCN detections requires either densities $\geq10^5$~cm$^{-3}$, or a substantial amount of mechanical heating. Future observations of dense-gas tracer excitation ladders in individual DSFGs will provide further constraints on the role of turbulence for the energetics of DSFGs.

\subsection{What causes the low HCN/CO fraction in DSFGs?}
What is the source of the wide spread between high $f_\mathrm{dense}$ in SDP.9 and J16359, and the low $f_\mathrm{dense}$ in our sample?

We consider the four following scenarios:
\begin{itemize}
    \item Dense gas is depleted in DSFGs with strong radiation fields (e.g. \citealt{Hopkins2013}).
    \item Dense gas content is enhanced in DSFGs undergoing major mergers (e.g. \citealt{Juneau2009, Bournaud2010}).
  	\item The HCN emission in some DSFGs is boosted by compact gas reservoir size and/or AGNs (e.g. \citealt{Privon2015}).
    \item The HCN-to-dense gas conversion factor $\alpha_\mathrm{HCN}$ varies significantly ($\sim$1~dex) between individual galaxies (e.g. \citealt{Vollmer2017}).
\end{itemize}

The following discussion has to be taken with a grain of salt, as the quality of data available for individual DSFGs varies considerably: from the exquisite 100-pc dust, CO, and [\ion{C}{ii}] imaging of SDP.81 \citep{alma2015, Rybak2015a, Dye2015, Rybak2020b}, to low-fidelity, marginally spatially resolved imaging of J16359 \citep{Kneib2005, Weiss2005}. Future high-resolution, multi-tracer observations will be necessary to characterise the kinematic status and physical conditions in individual sources.

\begin{figure}
\begin{centering}
\includegraphics[height=5cm, clip=true]{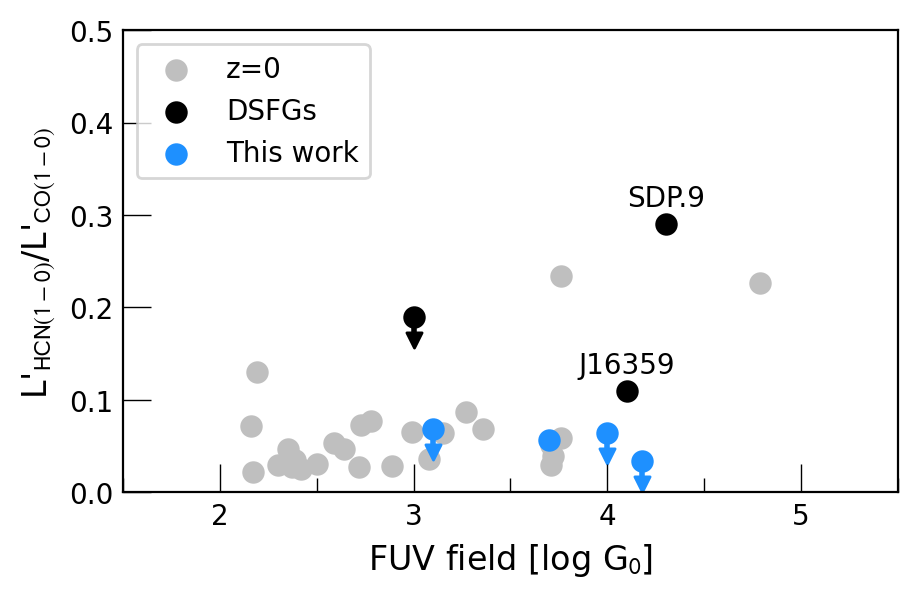}\\
\includegraphics[height=5cm, clip=true]{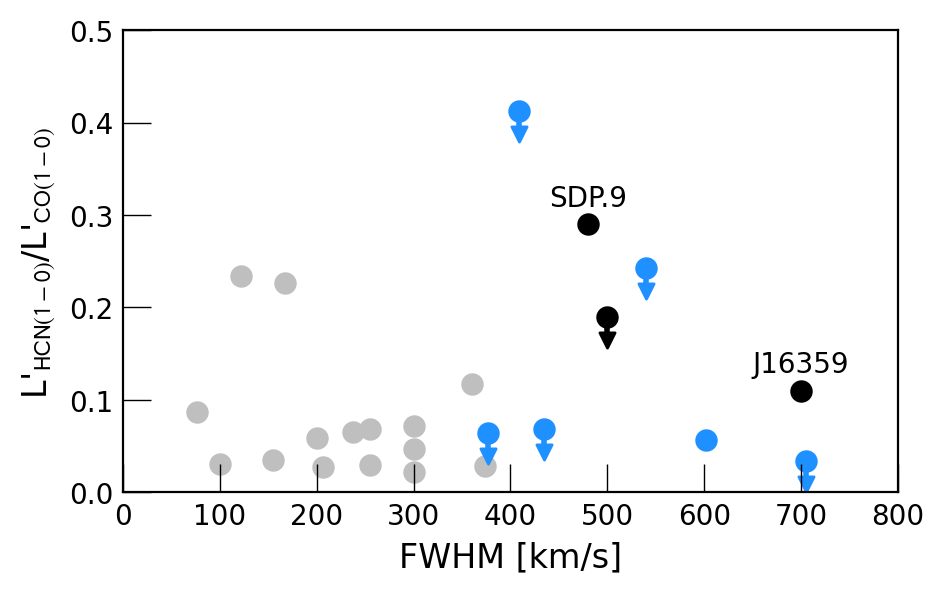}\\
\includegraphics[height=5cm, clip=true]{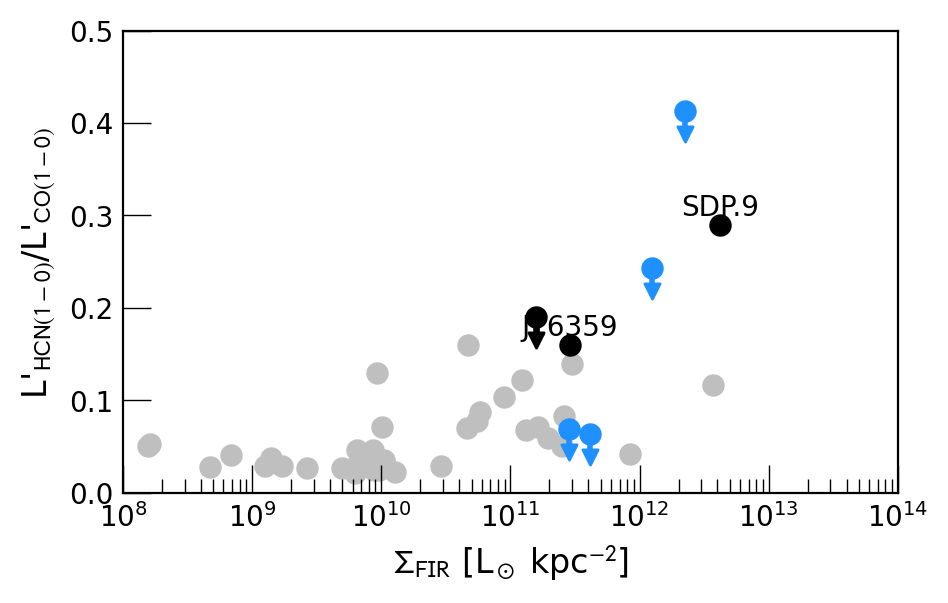}\\
\caption{HCN/CO ratios vs FUV field strength (upper) and CO FWHM (center) and $\Sigma_\mathrm{FIR}$ in our sample and literature data \citep{Garcia2012, Diaz2017}. The FUV field is inferred using \textsc{PDRToolbox}. There is no clear trend between the radiation field strength or CO line FWHM and the HCN/CO ratio. We find a tentative trend in increasing HCN/CO ratio with $\Sigma_\mathrm{FIR}$, potentially boosted by an X-ray bright AGN in SDP.9.}
\label{fig:HCN_CO_tests}
\end{centering}
\end{figure}

\subsubsection{Dense gas depleted by radiation field}
\label{subsubsec:HCN_CO_FUV}

Given the high $\Sigma_\mathrm{SFR}$ in DSFGs, their molecular clouds will be exposed to intense FUV fields from nearby young stars, which will alter their density distribution by photoevaporation and radiation pressure. Using zoom-in simulations of individual galaxies, \citet{Hopkins2013} found that the HCN-emitting, high-density gas is very sensitive to the radiation pressure, whereas the bulk of CO-emitting gas is essentially unaffected.

To test this scenario, we compare the FUV radiation field and HCN/CO ratio in both high-$z$ and nearby galaxies. As a first-order approximation, assuming the same intrinsic stellar SED, the FUV field strength and radiation pressure should scale with each other. We adopt the source-averaged FUV field strength $G$ in $G_0$ units\footnote{1~$G_0=1.6\times10^{-3}$ erg cm$^{-2}$ s$^{-1}$, integrated between 91.2 and 240~nm.} from the literature \citep{Canameras2018,Rybak2020b}. For comparison, we include 28 $z\sim0$ sources from the \citet{Gao2007} and \citet{Garcia2012} samples, which have [\ion{C}{ii}] 158-$\mu$m, [\ion{O}{i}] 63-$\mu$m, and FIR continuum measurements from \citet{Diaz2017}. The FUV field strengths were derived using the \textsc{PDRToolbox} models \citep{Kaufman1999,Kaufman2006,Pound2008}, assuming a double-side illumination, and that [\ion{C}{ii}] and dust continuum are optically thin. 

As shown in Fig~\ref{fig:HCN_CO_tests}, there is no clear trend in HCN/CO as a function of $G$. For example, J1202 has $G$ about $\sim$0.5~dex lower than SDP.9, yet its HCN/CO ratio is 10$\times$ lower. Similarly, SDP.81 has the lowest source-averaged $G$ from the entire sample, yet it has HCN/CO$\leq$0.07. This does not rule out that the FUV fields in the star-forming regions are much higher than the source-averaged values, as seen in the recent study of SDP.81, where the FUV field varies by $\geq$1~dex on sub-kpc scales \citep{Rybak2020b}.

\subsubsection{Merger-driven density enhancement}

Major mergers of gas-rich galaxies might enhance the high-density tail of the gas distribution function \citep{Bournaud2010, Powell2013, Moreno2019}. Such enhancements may occur already in the early stages in the merger at $\geq$10~kpc separations \citep{Sparre2016}, rather than only during coalescence. Indeed, HCN/CO ratios have been proposed as a tool to distinguish between disk and merger/nuclear modes of star-formation \citep{Papadopoulos2012}.

In this scenario, J16359 and SDP.9 should show more prominent merger signatures than our sample. Due to the lack of high-resolution kinematic maps, we focus on the source-averaged low-$J$ CO spectra: our proxies for mergers are the CO linewidth and whether the spectrum shows multiple peaks \citep{Bothwell2013}.

Figure~\ref{fig:HCN_CO_tests} shows the HCN/CO ratio as a function of the CO FWHM for our sample and the $z\sim0$ sources from \citet{Gao2004b} and \citet{Garcia2012}. Encouragingly, J16359 shows a prominent double-peaked profile in CO(3--2) and higher-J lines \citep[IRAM]{Weiss2005} with a peak separation of $\sim$300 km s$^{-1}$, indicative of a major merger. However, such prominent features are absent in the SDP.9 CO(3--2) spectrum \citep{Iono2012, Oteo2017}, although this source has the highest HCN/CO ratio out of all DSFGs. Two more sources -- HXMM.02 and J1609 show a double-peaked profile and a large FWHM: while the HCN/CO limit for HXMM.02 is very weak, J1609 has HCN/CO$\leq0.034$, 3$\times$ lower than J16359. SDP.81, SDP.130, and J1202 do not exhibit a double-peak line profile (although SDP.81 spectrum becomes double-peaked after de-lensing \citep{Rybak2015b}. We therefore do not find any clear correlation in HCN/CO ratio with the merger stage.

\subsubsection{FIR size and AGN contribution}
\label{subsubsec:HCN_CO_AGN}

The final plot in Fig.~\ref{fig:HCN_CO_tests} shows the HCN/CO ratio as a function of the mean star-formation rate surface density $\langle\Sigma_\mathrm{FIR}\rangle$, where $\langle\Sigma_\mathrm{FIR}\rangle=L_\mathrm{FIR}/2/(\pi r_{1/2}^2)$. For the lensed DSFGs, we adopt FIR half-light radius ($r_\mathrm{1/2}$) measurements from \citet[J16359]{Kneib2005}, \citet[HXMM.02, SDP.11]{Bussmann2013,Bussmann2015}, \citet[SDP.9]{Massardi2018}, \citet[SDP.81]{Rybak2015a}, \citet[SDP.130]{Falgarone2017} and \citet[J0209]{Geach2018}. For the $z\sim0$ sources, we use the 70-$\mu$m sizes from \citet{Lutz2016}. 

Figure~\ref{fig:HCN_CO_tests} indicates a qualitative trend of HCN/CO ratio increasing with $\Sigma_\mathrm{FIR}$.
Namely, SDP.9 - the source with the highest HCN/CO ratio - has a very compact FIR morphology ($r_{1/2}\simeq$300~pc), with $\Sigma_\mathrm{FIR}$ almost 1~dex higher than any other source in our sample. Moreover, SDP.9 and SDP.11 are detected in the \textit{Chandra} X-ray imaging, indicating a powerful buried AGN \citep{Massardi2018}; SDP.9 also shows an unusually excited CO SLED, which is more consistent with AGN hosts than purely star-forming galaxies \citep{Oteo2017}. Indeed, X-ray emission from AGNs can significantly boost the HCN or HCO$^+$ (e.g. \citealt{Privon2020}); such effects will be particularly significant for compact source sizes. It is therefore plausible that the large HCN/CO ratio in SDP.9 is driven by the buried AGN rather than by its high dense-gas fraction.

Unfortunately, out of our entire sample, only SDP.81 is detected in the XMM-Newton imaging ($L_\mathrm{0.5-4.5~keV}\simeq49\times10^{-15}$ erg s$^{-1}$ cm$^{-2}$; \citealt{Ranalli2015}), while HXMM.02 has informative upper limits ($L_\mathrm{0.5-4.5~keV}\leq8\times10^{-16}$ erg s$^{-1}$ cm$^{-2}$;  \citealt{Ueda2008,Ikarashi2011}); the remaining sources do not have X-ray data of sufficient quality to allow a proper comparison.

\subsubsection{Environmental dependence of $\alpha_\mathrm{HCN}$}

Finally, we note that the wide spread in HCN/CO ratios in DSFGs might be due to a varying $\alpha_\mathrm{HCN}\equiv M_\mathrm{dense} / L'_\mathrm{HCN}$ - the conversion factor between $L'_\mathrm{HCN}$ and dense gas mass. 
This problem parallels the well-known uncertainty in the CO-to-$M_\mathrm{gas}$ conversion factor - $\alpha_\mathrm{CO}$ - which shows strong environmental dependence, dropping from $\alpha_\mathrm{CO}=4.4$~$M_\odot$/(K km s$^{-1}$ pc$^2$) in the Milky Way to $\alpha_\mathrm{CO}\sim1$~$M_\odot$/(K km s$^{-1}$ pc$^2$) in ULIRGs and DSFGs, though estimates for individual galaxies vary widely.

For the HCN, \citet{Gao2004b} propose $\alpha_\mathrm{HCN}=10$~$M_\odot$/(K km s$^{-1}$ pc$^2$). Although this value is widely quoted, it would imply dense-gas fractions greater than 100\% for ULIRGs and DSFGs which have see HCN/CO$\geq$0.1 (see Fig.~\ref{fig:CO_HCN}, which is clearly unphysical. The value of $\alpha_\mathrm{HCN}$ in intensely star-forming galaxies thus remains highly controversial.

On the observational side, \citet{Gracia2008} used large velocity gradient (LVG) modelling of HCN and HCO$^+$ (1--0) and (3--2) emission in nearby (U)LIRGS to argue for $\alpha_\mathrm{HCN}\leq2.5$ in ULIRGs. On the other hand, several recent simulations on cloud- and galaxy-scales predict elevated $\alpha_\mathrm{HCN}$; e.g. \citet{Onus2018} predict $\alpha_\mathrm{HCN}=14\pm6$ $M_\odot$/(K km s$^{-1}$ pc$^2$) and \citet{Vollmer2017} report $\alpha_\mathrm{HCN}=33\pm17$ $M_\odot$/(K km s$^{-1}$ pc$^2$) for DSFG-like galaxies. A higher $\alpha_\mathrm{HCN}$ would make dense gas fainter in HCN emission, allowing DSFGs to have high dense-gas fractions while retaining low HCN/CO luminosity ratios. 

Given the lack of appropriate constraints, we choose to remain agnostic with respect to the actual value of $\alpha_\mathrm{HCN}$ in DSFGs. Following the approach of \citet{Gracia2008}, future observations of HCN and HCO$^+$ ladders in DSFGs are required to better constrain $\alpha_\mathrm{HCN}$ in high-z galaxies.

\section{Conclusions}
\label{sec:conclusions}

We have conducted the largest survey of dense gas content in high-redshift DSFGs, by targeting the $J=1-0$ HCN, HCO$^+$ and HNC emission in six $z\sim3$ strongly lensed DSFGs from the H-ATLAS and \textit{Planck} samples. Our main findings are:

\begin{itemize}
    \item We detect the HCN(1--0) emission in J1202; HCO$^+$(1--0) is tentatively detected in J1609. No other HCN, HCO$^+$ or HNC lines are detected, either in individual sources. Stacking the spectra of the six sources does not yield any detections.
    \item We find HCN/FIR ratios $\leq(4-10)\times10^{-4}$ for individual galaxies and $\leq3.4\times10^{-4}$ for the stacked spectrum. These are somewhat lower than the general HCN-FIR linear correlation \citep{Bigiel2016, Jimenez2019}, but consistent with $z\sim0$ ULIRGs and previous high-$z$ detections \citep{Gao2007, Oteo2017}. The departure from a linear $L'_{HCN}$--$L_\mathrm{FIR}$ trend corroborates theoretical arguments \citep{Krumholz2007, Narayanan2008} that predict a sub-linear trend at high SFRs. 
    \item Using the \citet{Onus2018} relation between HCN/FIR ratio and $\epsilon_\mathrm{ff}$, we find a source-averaged star-formation efficiency per free-fall time $\epsilon_\mathrm{ff}=$1\% (J1202) and $\geq$2\% (stack), consistent with dynamical arguments for high-$z$ starbursts.
    \item From the stacked spectrum, we find HCN/CO ratio $\leq0.045$; the HCN(1--0) detection in J1202 similarly implies HCN/CO$\approx$0.05. These are similar to $z\sim0$ sub-ULIRG galaxies with $L_\mathrm{FIR}\leq10^{11}$~$L_\odot$, that is, significantly lower than in many ULIRGs and the two previously reported DSFG detections.
    Contrary to the previous reports, \emph{the majority of DSFGs have a relatively low dense-gas fraction and somewhat elevated star-formation efficiency}.
    
    \item A comparison with HCN/CO/FIR predictions from \citet{Krumholz2007} shows that the DSFGs are more akin to ''normal'' star-forming galaxies rather than starbursts. The inferred mean gas density in DSFGs is lower than $10^3$ cm$^{-3}$. We hypothesise that the bulk of CO/[\ion{C}{ii}]-traced molecular gas is in extended reservoirs and not directly associated with star formation, as suggested by high-resolution imaging of individual DSFGs (e.g. \citealt{Riechers2011, Hodge2015, Calistro2018,Rybak2020b})
    
    \item Comparing the current HCN $J_\mathrm{upp}=1-5$ detections in DSFGs to radiative transfer models that include mechanical heating \citep{Kazandjian2015}, we find that reproducing the current data requires either very high gas densities ($\geq10^5$~cm$^{-3}$), or a substantial mechanical heating. However, obtaining fully-sampled HCN ladders in high-z galaxies is paramount for properly characterise the physical conditions in the dense-gas phase.

    \item We consider different mechanisms  (e.g. $\Sigma_\mathrm{SFR}$, radiation pressure or major mergers) that might cause the wide spread between the elevated HCN/CO ratios reported for SDP.9 and J16359, and the low HCN/CO ratios in our sample. We find a tentative trend in HCN/CO increasing with $\Sigma_\mathrm{SFR}$, and hypothesise a buried AGN drives the atypically high HCN/CO ratio in SDP.9.

\end{itemize}

The faintness of the HCN, HCO$^+$, and HNC(1--0) emission in our highly magnified sources highlights the challenges in studying dense gas at high redshift with current facilities. Large-scale surveys of dense-gas tracers in high-$z$ galaxies will remain challenging even with the planned Next Generation Very Large Array. Consequently, strong gravitational lensing remains an indispensable tool for studying the cold, high-density ISM in high-redshift galaxies.

\begin{acknowledgements}

We thank Elisabete da Cunha, Mark Krumholz, and Desika Narayanan for helpful discussions and valuable insights from simulations and theory.

The National Radio Astronomy Observatory is a facility of the National Science Foundation operated under cooperative agreement by Associated Universities, Inc.
M. R. is supported by the NWO Veni project "Under the lens" (VI.Veni.202.225). M. R. and J. H. acknowledge support of the VIDI research programme with project number 639.042.611, which is (partly) financed by the Netherlands Organisation for Scientific Research (NWO).
I.L. acknowledges support from the Comunidad de Madrid through the Atracci\'on de Talento Investigador Grant 2018-T1/TIC-11035.

\end{acknowledgements}

%-------------------------------------------------------------------
\bibliographystyle{aa}
\bibliography{HCN_survey_2lines.bbl}

%-------------------------------------------------------------------
\begin{appendix}

\section{Limits on the SiO(2--1) and CS(2--1) line luminosities}
\label{app:A}

In addition to the HCN, HCO$^+$ and HNC(1--0) lines discussed in this Paper, the spectral setup of our JVLA observations covers the SiO(2--1) line ($f_0=86.847$~GHz)  in all six galaxies, and the CS(2--1) line ($f_0=97.981$~GHz) in SDP.81 and HXMM.02. None of these lines are detected at $\geq3\sigma$ level - neither in the individual spectral (Fig.~\ref{fig:vla_spectra}), the stacked spectrum (Fig.~\ref{fig:vla_stacks}) or narrow-band imaging. Table~\ref{tab:sio_cs} lists the 3-$\sigma$ upper limits for individual galaxies and the stacked spectrum. Our spectral stacking implies 3$\sigma$ upper limits on $L'_\mathrm{SiO(2-1)}/L_\mathrm{FIR}\leq4.2\times10^{-4}$ and $L'_\mathrm{CS(2-1)}/L_\mathrm{FIR}\leq4.0\times10^{-4}$.

\begin{table*}
    \caption{Upper limits (3-$\sigma$) on the SiO(2--1) and CS(2--1) line luminosities for individual galaxies and the stacked spectrum ($L_\mathrm{FIR}=5\times10^{12}$~$L_\odot$). \label{tab:sio_cs}}
    \label{tab:cs_sio}
    \centering
    \begin{tabular}{l|cccc}
    \hline
     Galaxy &    $L^{\mathrm{sky}}_\mathrm{SiO(2-1)}$ &  $L'^{\mathrm{sky}}_\mathrm{SiO(2-1)}$ &  $L^{\mathrm{sky}}_\mathrm{CS(2-1)}$ &  $L'^{\mathrm{sky}}_\mathrm{CS(2-1)}$ \\
     & [$L_\odot$] & [K km s$^{-1}$ pc$^2$]& [$L_\odot$] & [K km s$^{-1}$ pc$^2$] \\
     \hline
     SDP.81 & $\leq6.3\times10^{5}$ & $\leq3.3\times10^{10}$ &  $\leq6.0\times10^{5}$ & $\leq2.0\times10^{10}$\\
     SDP.130 & $\leq2.8\times10^{5}$ & $\leq1.6\times10^{10}$ & --- & ---\\
     HXMM.02 & $\leq17.3\times10^{5}$ & $\leq41.4\times10^{10}$ &  $\leq12\times10^{5}$ & $\leq3.0\times10^{10}$ \\
     J0209 & $\leq54.4\times10^{5}$ & $\leq20.2\times10^{10}$ & --- & ---\\
     J1202 & $\leq30.5\times10^{5}$ & $\leq9.7\times10^{10}$ &  --- & --- \\
     J1609 & $\leq23.6\times10^{5}$ & $\leq41.2\times10^{10}$ & --- & --- \\
     \hline
     Stack & $\leq5.5\times10^{4}$ & $\leq2.1\times10^{9}$ & $\leq3.8\times10^{4}$ & $\leq2.0\times10^{9}$  \\
     \hline
    \end{tabular}
    
\end{table*}

\section{Attenuation of dense-gas tracers due to the cosmic microwave background}
\label{app:B}

The temperature of the cosmic microwave background (CMB) increases with redshift as $T_\mathrm{CMB}=(1+z)\times2.73$~K. Although negligible at present-day, at high redshift, the CMB continuum provides additional source of heating and a background against which any emission line is observed. These two effects can significantly attenuate the observed emission from the cold interstellar medium \citep{daCunha2013,Zhang2016}. Can the faint HCN emission in our $z\sim3$ sample be explained by attenuation due to the CMB background?

First, we follow \citet{daCunha2013} by calculating the expected attenuation due to the elevated CMB temperature for each galaxy, assuming that the dense gas and dust are in a local thermal equilibrium (i.e. excitation temperature and dust temperature are balanced, $T_\mathrm{exc}=T_\mathrm{dust}$). In this approximation, the line flux density observed against CMB at frequency $\nu$ is given by:

\begin{equation}
S_\mathrm{obs}(\nu)=\frac{\Omega}{(1+z)^3} \Big(1-\exp(-\tau_nu_0)\Big)  \Big( B(T_\mathrm{exc}, \nu_0) - B(T_\mathrm{CMB}(z), \nu_0) \Big),
\label{eq:daCunha}
\end{equation}

where $\Omega$ is the solid angle subtended by the galaxy, $\tau_nu_0$ is the optical depth of the line transition, and $B(T, \nu)$ is the black-body fuction for temperature $T$ evaluated at a rest-frame frequency $\nu_0=\nu(1+z)$.

For each source, we adopt the $T_\mathrm{dust}$ derived from far-IR and mm-wave photometry \citep{Bussmann2015, Rybak2019, Harrington2021}; these range between 35~K and 42~K. The corresponding attenuation due to the CMB ranges between 20 - 30\%. 

However, the local thermal equilibrium approximation can not be easily extended to the scales of entire galaxies, which might exhibit significant temperature, density and optical depth gradients on kpc-scales. The impact of the CMB temperature on high-z observations of HCN, CO, and FIR continuum in this more complex scenario was explored by \citet{Tunnard2017}. in particular, they used both single-line-of-sight and galaxy-scale toy models coupled with the \textsc{Radex} \citep{VanDerTak2007} radiative transfer calculations. In the latter, using a toy model of NGC~1068, they find that HCN(1--0) is attenuated by $\leq$10\% at $z=2.5-3.5$; in an extreme case of a ``cool'' NGC~1068 model (kinetic temperature at the outskirts lowered from 40~K to 15~K), HCN(1--0) is attenuated by up to 30\%. These results are broadly in line with the predictions of Eq.~\ref{eq:daCunha}. 

Consequently, we do not consider CMB to be the cause of the low observed HCN fluxes of our sample. We do not apply any additional correction to the line luminosities in Tab.~\ref{tab:vla_results}. As the CMB effect on the FIR continuum at $z\sim3$ is also negligible, we consider the observed HCN/FIR ratios to be robust.

Considering the HCN/CO ratio, even though the CO(1--0) emission is more susceptible to the CMB temperature, \citet{Tunnard2017} do not find that the observed HCN/CO ratios at $z=3$ are significantly overestimated. However, their models consider relatively compact cold gas reservoir sizes (scale radius 1.4~kpc); CO(1--0) emission from very extended ($\geq10$~kpc) cold gas reservoirs seen in some DSFGs might be suppressed more efficiently \citep{daCunha2013,Zhang2016}. In such a case, the intrinsic HCN/CO ratios reported in Tab.~\ref{tab:vla_results} would decrease further.

\end{appendix}

\end{document}